\documentclass[fleqn,usenatbib]{mnras}

\usepackage{graphicx}	% Including Fig.files
\usepackage{amsmath}	% Advanced maths commands
\usepackage{amssymb}	% Extra maths symbols
\usepackage{hyperref} 
\usepackage{longtable}
\usepackage{supertabular} 
\usepackage{booktabs}
\usepackage{placeins}
\usepackage{lscape}
\usepackage{xspace} % fixes spacing issue with \jbu.. commands
\usepackage{multicol}
\usepackage{float}
\usepackage{siunitx}
\usepackage{comment}
\usepackage{enumitem}
\usepackage{upgreek}

\usepackage{tablefootnote}
\graphicspath{{./Figures/}}

\newcommand{\orcid}[1]{\href{https://orcid.org/#1}{\includegraphics[width=10pt]{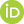}}}

\newcommand{\jbu}{AT~2016jbu\xspace}
\newcommand{\etacar}{$\eta$ Car\xspace}
\newcommand{\irc}{IRC+10420\xspace}
\newcommand{\ip}{SN~2009ip\xspace}
\newcommand{\gc}{SN~2013gc\xspace}
\newcommand{\bdu}{SN~2016bdu\xspace}
\newcommand{\bh}{SN~2015bh\xspace}
\newcommand{\lsq}{LSQ13zm\xspace}
\newcommand{\al}{SN~1996al\xspace}

\newcommand{\msun}{M$_\odot$}
\newcommand{\msunperyr}{$\mathrm{M}_\odot~\mathrm{yr}^{-1}$}
\newcommand{\kms}{km~s$^{-1}$}

\newcommand{\eventA}{\textit{Event A}\xspace}

\newcommand{\eventB}{\textit{Event B}\xspace}

\newcommand{\paperI}{\textcolor{blue}{Paper I}\xspace}

\title[Progenitor, environment, and modelling of the interacting transient, \jbu ]{Progenitor, environment, and modelling of the interacting transient, \jbu (Gaia16cfr)}

\author[S. J. Brennan et al.]{S. J. Brennan$^{1}$\orcid{0000-0003-1325-6235}%
\thanks{Contact e-mail: \href{sean.brennan2@ucdconnect.ie}{sean.brennan2@ucdconnect.ie}},
M. Fraser$^{1}$\orcid{0000-0003-2191-1674},
J. Johansson$^{2}$\orcid{0000-0001-5975-290X},
A. Pastorello$^{3}$\orcid{0000-0002-7259-4624},
R. Kotak$^{4}$\newauthor 
H. F. Stevance$^{5}$,
T. -W. Chen$^{6,7}$\orcid{0000-0002-1066-6098},
J. J. Eldridge$^{5}$\orcid{0000-0002-1722-6343},
S. Bose$^{8,9}$,
P. J. Brown$^{10}$\orcid{0000-0001-6272-5507},
E. Callis$^{1}$\orcid{0000-0002-1178-2859},\newauthor 
R. Cartier$^{11}$,
M. Dennefeld$^{12}$,
Subo Dong$^{13}$,
P. Duffy$^{1}$,
N. Elias-Rosa$^{14,15}$\orcid{0000-0002-1381-9125},
G. Hosseinzadeh$^{16}$\orcid{0000-0002-0832-2974},\newauthor 
E. Hsiao$^{17}$\orcid{0000-0003-1039-2928},
H. Kuncarayakti$^{18,19}$,
A. Martin-Carrillo$^{1}$,
B. Monard$^{20}$,
G. Pignata$^{21,22}$\orcid{0000-0003-0006-0188},\newauthor 
D. Sand$^{23}$\orcid{0000-0003-4102-380X},
B. J. Shappee$^{24}$\orcid{0000-0003-4631-1149},
S. J. Smartt$^{25}$,
B. E. Tucker$^{26,27,28}$\orcid{0000-0002-4283-5159},
L. Wyrzykowski$^{29}$\orcid{0000-0002-9658-6151},\newauthor 
H. Abbot$^{26}$,
S. Benetti$^{3}$\orcid{0000-0002-3256-0016},
J. Bento$^{26}$, 
S. Blondin$^{30,31}$\orcid{0000-0002-9388-2932},
Ping Chen$^{28}$,
A. Delgado$^{32,33}$,
L. Galbany$^{34}$\orcid{0000-0002-1296-6887},\newauthor
M. Gromadzki$^{29}$\orcid{0000-0002-1650-1518},
C. P. Guti\'errez$^{19,35}$\orcid{0000-0003-2375-2064},
L. Hanlon$^{1}$,
D. L. Harrison$^{32,36}$\orcid{0000-0001-8687-6588},
D. Hiramatsu$^{37,38,54,55}$\orcid{0000-0002-1125-9187},\newauthor 
S. T. Hodgkin$^{32}$\orcid{0000-0002-5470-3962},
T. W.-S. Holoien$^{39}$\orcid{000-0001-9206-3460},
D. A. Howell$^{37,38}$\orcid{0000-0003-4253-656X},
C. Inserra$^{40}$\orcid{0000-0002-3968-4409},
E. Kankare$^{4}$\orcid{0000-0001-8257-3512},\newauthor
S. Koz{\l}owski$^{29}$\orcid{0000-0003-4084-880X}, 
T. E. M\"{u}ller-Bravo$^{41,53}$\orcid{0000-0003-3939-7167},
K. Maguire$^{42}$\orcid{0000-0002-9770-3508},
C. McCully$^{37,38}$\orcid{0000-0001-5807-7893},
P. Meintjes$^{43}$,\newauthor 
N. Morrell$^{44}$\orcid{0000-0003-2535-3091},
M. Nicholl$^{45,46}$,
D. O'Neill$^{25}$,
P. Pietrukowicz$^{29}$\orcid{0000-0002-2339-5899},
R. Poleski$^{29}$\orcid{0000-0002-9245-6368},
J. L. Prieto$^{22,47}$,\newauthor 
A. Rau$^{7}$,
D. E. Reichart$^{48}$\orcid{0000-0002-5060-3673},
T. Schweyer$^{6,7}$,
M. Shahbandeh$^{49}$,
J. Skowron$^{29}$\orcid{0000-0002-2335-1730}, 
J. Sollerman$^{6}$\orcid{0000-0003-1546-6615},\newauthor
I. Soszy{\'n}ski$^{29}$\orcid{ 0000-0002-7777-0842},
M. D. Stritzinger$^{50}$\orcid{0000-0002-5571-1833},
M. Szyma{\'n}ski$^{29}$\orcid{ 0000-0002-0548-8995},
L. Tartaglia$^{3}$\orcid{0000-0003-3433-1492},
A. Udalski$^{29}$\orcid{0000-0001-5207-5619},\newauthor 
K. Ulaczyk$^{29,51}$\orcid{0000-0001-6364-408X},
D. R. Young$^{52}$\orcid{0000-0002-1229-2499},
M. van Leeuwen$^{32}$,
B. van Soelen$^{43}$
\\
The authors' affiliations are shown in Appendix \ref{app:affiliations}.
}

\pubyear{2021}

\begin{document}

\label{firstpage}
\pagerange{\pageref{firstpage}--\pageref{lastpage}}
\maketitle

\begin{abstract}
% If we need to shorten the abstract then cut down the first sentence. The photometric variability is quite important!
We present the bolometric lightcurve, identification and analysis of the progenitor candidate, and preliminary modelling of \jbu (Gaia16cfr). We find a progenitor consistent with a $\sim$~22--25~\msun\ yellow hypergiant surrounded by a dusty circumstellar shell, in agreement with what has been previously reported. We see evidence for significant photometric variability in the progenitor, as well as strong H$\alpha$ emission consistent with pre-existing circumstellar material. The age of the environment as well as the resolved stellar population surrounding \jbu, support a progenitor age of $>$10 Myr, consistent with a progenitor mass of $\sim$22~\msun. A joint analysis of the velocity evolution of \jbu, and the photospheric radius inferred from the bolometric lightcurve shows the transient is consistent with two successive outbursts/explosions. The first outburst ejected material with velocity $\sim$650~\kms, while the second, more energetic event, ejected material at $\sim$4500~\kms. Whether the latter is the core-collapse of the progenitor remains uncertain. We place a limit on the ejected $^{56}$Ni mass of $<$0.016\msun. Using the {\sc BPASS} code, we explore a wide range of possible progenitor systems, and find that the majority of these are in binaries, some of which are undergoing mass transfer or common envelope evolution immediately prior to explosion. Finally, we use the {\sc SNEC} code to demonstrate that the low-energy explosion within some of these binary systems, together with sufficient CSM, can reproduce the overall morphology of the lightcurve of \jbu.
\end{abstract}

\begin{keywords}
 circumstellar matter – stars: massive – supernovae: general – supernovae: individual: \jbu 
\end{keywords}
%%%%%%%%%%%%%%%%%%%%%%%%%%%%%%%%%%%%%%%%%%%%%%%%%%

% Shorthand for kilpatrick et al. 2018
\defcitealias{Kilpatrick2018}{K18}

%%%%%%%%%%%%%%%%% BODY OF PAPER %%%%%%%%%%%%%%%%%%

\section{Introduction}\label{sec:intro}

This is the second of two papers on the interacting transient \jbu (Gaia16cfr). We report photometric and spectroscopic observations in \citealt{brennan2021a} (hereafter \paperI) and present an in-depth comparison of \jbu and \ip-like transients which include \ip \citep{Fraser13, Graham2014}, \bh \citep{Elias-Rosa2016,Thone2017}, \lsq \citep{Tartaglia2016}, \gc \citep{Reguitti2018} and \bdu \cite{Pastorello_2017}. The work presented here will focus on the progenitor candidate, its environment as well as modelling and interpretation of the spectral and photometric evolution.

\jbu shows a smooth evolution of the H$\alpha$ emission profile, changing from a P Cygni profile, typically seen in Type II supernova (SN) spectra which show strong, singular peaked, hydrogen emission lines \citep{Kiewe2012,Taddia_2015}, to a double-peaked emission profile which persists until late times, indicating complex, H-rich, circumstellar material (CSM). \jbu and \ip-like objects show strong similarities in late time spectra with strong \ion{Ca}{II}, \ion{He}{I} and H emission lines as well as a lack of any emission from explosively nucleosynthesised material such as $[\ion{O}{I}]~\lambda\lambda~6300,6364$ or $\ion{Mg}{I}]~\lambda4571$. No clear nebular phase is seen even after $\sim$~1.5~years after explosion in \jbu, and on-going interaction with CSM at late times may be hiding a nebular phase and/or inner material from the progenitor.

The nature of \ip-like transients is much more contentious. On one hand, there is evidence that these are genuine core-collapse supernovae (CCSNe), the progenitor was destroyed and the transient will fade after CSM interaction finishes \citep{Pastorello2013,Smith2014,Pastorello2019,Graham2014,atel4412}. On the other hand, some suggest these may be non-terminal events \citep{Fraser13,Margutti2014,Fraser2015,Graham2017}, and \ip-like events are a result of either pulsational-pair instabilities \citep{Woosley_2007,Woosley_2017}, binary interaction \citep{Kashi2013,Pastorello2019}, merging of massive stars \citep{Soker_2013} or instabilities associated with rapid rotation close to the $\Omega\Gamma$-limit \citep{Maeder_2000}.

As a follow-up to \paperI we continue the discussion on \jbu, focusing on the progenitor and its local environment, as well as examining the controversial topic of the powering mechanism behind \ip-like events. We note that some of these topics have been discussed before by \citealp{Kilpatrick2018} (hereafter referred to as \citetalias{Kilpatrick2018}), and we refer to this work throughout. For consistency with \paperI, and to compare to previous work by \citetalias{Kilpatrick2018}, we take the distance modulus for NGC~2442 to be $31.60\pm0.06$~mag. This corresponds to a distance of $20.9\pm0.58$~Mpc and adopt a redshift z=0.00489 from the H I Parkes All Sky Survey (HIPASS) \citep{Wong2006}. The foreground extinction towards NGC~2442 is taken to be $A_V=0.556$~mag \citep{Schlafly11} via the NASA Extragalactic Database (NED;\footnote{\url{https://ned.ipac.caltech.edu/}}). We correct for foreground extinction using $R_V=3.1$ and the extinction law given by \citealp{Cardelli1989}. We do not correct for any host galaxy or circumstellar extinction, however note that the blue colours seen in the spectra of \jbu do not point towards significant reddening by additional dust (further discussed in Sect.~\ref{sec:muse} and Sect.~\ref{sec:progenitor}). We take the {\it V}-band maximum at \eventB (as determined through a polynomial fit) as our reference epoch (MJD $57784.4\pm0.5$; 2017 Jan 30). Significant lightcurve features will use the same naming convention as in \paperI for specific points in the lightcurve; \textit{Rise, Decline, Plateau, Knee, Ankle}.

In Sect.~\ref{sec:bolometric_evolution} we investigate the CSM environment around \jbu and using photometry presented in \paperI reconstruct the bolometric evolution of \eventA and \eventB up until the seasonal gap ($+140$~days), which we discuss in Sect.~\ref{sec:kinematices}. The progenitor of \jbu is discussed in Sect.~\ref{sec:progenitor} using pre-explosion as well as late time imaging from the \textit{Hubble Space Telescope} ({\it HST}). This presence of pre-existing dust is discussed in \jbu using SED fitting as well as {\sc dusty} modelling in Sect.~\ref{sec:dust}. Using {\it HST} and Very Large Telescope (VLT) + Multi Unit Spectroscopic Explorer (MUSE) observations, we investigate the surrounding stellar population and environment in Sect.~\ref{sec:enviroment}. The powering mechanism behind \jbu is discussed in Sect.~\ref{sec:power}. In Sect.~\ref{sec:discuss_progenitor}, the most likely progenitor for \jbu is examined. \jbu and most \ip-like transients display a high degree of asymmetry, most likely due to a complex CSM environment, and this is expanded upon in Sect.~\ref{sec:discuss_geometry}. Finally, we will address the explosion scenario for \jbu and perhaps other \ip-like transients, focusing on a CCSN scenario in Sect.~\ref{sec:discuss_CCSN}, and an explosion in a binary system in Sect.~\ref{sec:discuss_binary}.

%%%%%%%%%%%%%%%%%%%%%%%%%%%%%%%%%%%
%%%%% Progenitor of AT~2016jbu %%%
%%%%%%%%%%%%%%%%%%%%%%%%%%%%%%%%%%%

\section{The progenitor of \jbu}\label{sec:progenitor}

The progenitor of \jbu was discussed by \citetalias{Kilpatrick2018}, who suggest that it was consistent with an F8 type star of $\sim$18~\msun\ from an optical SED fit, although circumstellar extinction places this as a lower bound. 

There is a wealth of pre-explosion images of NGC~2442 and in this section we explore this data to identify and characterise the progenitor of \jbu. Here we are specifically concerned with the quiescent (or apparently quiescent) progenitor which can only be identified in deep, high resolution data.

\subsection{Hubble Space Telescope imaging of the progenitor}\label{sec:hst}

\begin{table*}
\centering
\caption{Observational log for all {\it HST} images covering the site of \jbu. Measured photometry (in the Vega-mag system) for \jbu is also reported. Phase is in rest frame days relative to \eventB maximum light (MJD 57784.4) } 
\label{tab:hst}
\begin{tabular}{cccccc}
\hline
Date & Phase (d) & Instrument & Filter & Exposure (s) & Mag (err) \\
\hline
2006-10-20 & $-3736.0$ & ACS/WFC & F435W & 4$\times$395 & 24.999 (0.037) \\
- & - & - & F658N & 3$\times$450 & 21.207 (0.024) \\
- & - & - & F814W & 3$\times$400 & 23.447 (0.019) \smallskip\\
2016-01-21 & $-373.1$ & WFC3/UVIS & F350LP & 1$\times$420 & 23.625 (0.017) \\ 
- & - & WFC3/IR & F160W & 2$\times$503 & 20.726 (0.003) \\ 
2016-01-31 & $-362.9$ & WFC3/UVIS & F350LP & 2$\times$420 & 22.215 (0.026) \\ 
- & - & - & F555W & 2$\times$488 & 22.645 (0.002) \\ 
2016-02-08 & $-354.4$ & WFC3/UVIS & F350LP & 3$\times$420 & 22.134 (0.001) \\ 
- & - & WFC3/IR & F160W & 2$\times$503 & 19.570 (0.005) \\ 
2016-02-17 & $-345.5$ & WFC3/UVIS & F350LP & 3$\times$420 & 23.108 (0.012) \\ 
- & - & - & F814W & 2$\times$488 & 22.287 (0.003) \\ 
2016-02-23 & $-339.9$ & WFC3/UVIS & F350LP & 3$\times$420 & 23.212 (0.022) \\ 
2016-02-28 & $-334.8$ & WFC3/UVIS & F350LP & 1$\times$420 & 23.985 (0.022) \\ 
- & - & - & F555W & 2$\times$488 & 24.399 (0.004) \\ 
2016-03-04 & $-330.4$ & WFC3/UVIS & F350LP & 3$\times$420 & 22.729 (0.022) \\ 
- & - & WFC3/IR & F160W & 2$\times$503 & 20.224 (0.011) \\ 
2016-03-10 & $-323.8$ & WFC3/UVIS & F350LP & 3$\times$420 & 22.690 (0.037) \\ 
- & - & - & F814W & 2$\times$488 & 21.967 (0.022) \\ 
2016-03-15 & $-318.8$ & WFC3/UVIS & F350LP & 3$\times$420 & 22.868 (0.016) \\ 
- & - & WFC3/IR & F160W & 2$\times$503 & 20.323 (0.014) \\ 
2016-03-21 & $-313.1$ & WFC3/UVIS & F350LP & 3$\times$420 & 23.400 (0.013) \\ 
- & - & - & F555W & 2$\times$488 & 23.962 (0.012) \\ 
2016-03-30 & $-304.2$ & WFC3/UVIS & F350LP & 3$\times$420 & 23.775 (0.006) \\ 
- & - & WFC3/IR & F160W & 2$\times$503 & 21.301 (0.020) \\ 
2016-04-09 & $-293.7$ & WFC3/UVIS & F350LP & 3$\times$420 & 23.767 (0.006) \\ 
- & - & - & F814W & 1$\times$488 & 23.079 (0.035) \smallskip\\ 
2019-03-21 & +776.7 & WFC3/UVIS1 & F555W & 320,390 & 23.882 (0.025) \\ % MJD 58563.60 
- & - & - & F814W & 2$\times$390 & 23.239 (0.032) \smallskip\\
2019-03-31 & +787.0 & ACS/WFC & F814W & 4$\times$614 & 23.529 (0.014) \\ 
\hline
\end{tabular}
\end{table*}

NGC~2442 was observed with the \textit{Hubble Space Telescope} ({\it HST}) on a number of occasions both prior to and after the discovery of \jbu using the Advanced Camera for Surveys (ACS) and both the UV-Visible and IR channels of the Wide Field Camera 3 (WFC3/UVIS and WFC3/IR). We retrieved all images where the image footprint covered the site of \jbu from the Mikulski Archive for Space Telescopes (MAST\footnote{\url{mastweb.stsci.edu/}}), these data are listed in Table.~\ref{tab:hst}. In all cases, science-ready reduced images were downloaded. With the exception of the late-time ACS images taken in 2019, all analysis was performed on frames that have been already corrected for charge transfer efficiency losses at the pixel level (i.e. {\sc drc/flc} files). For the 2019 ACS images corrections for charge transfer efficiency were applied to the measured photometry.

\begin{figure*}
\centering
\includegraphics[height=10cm]{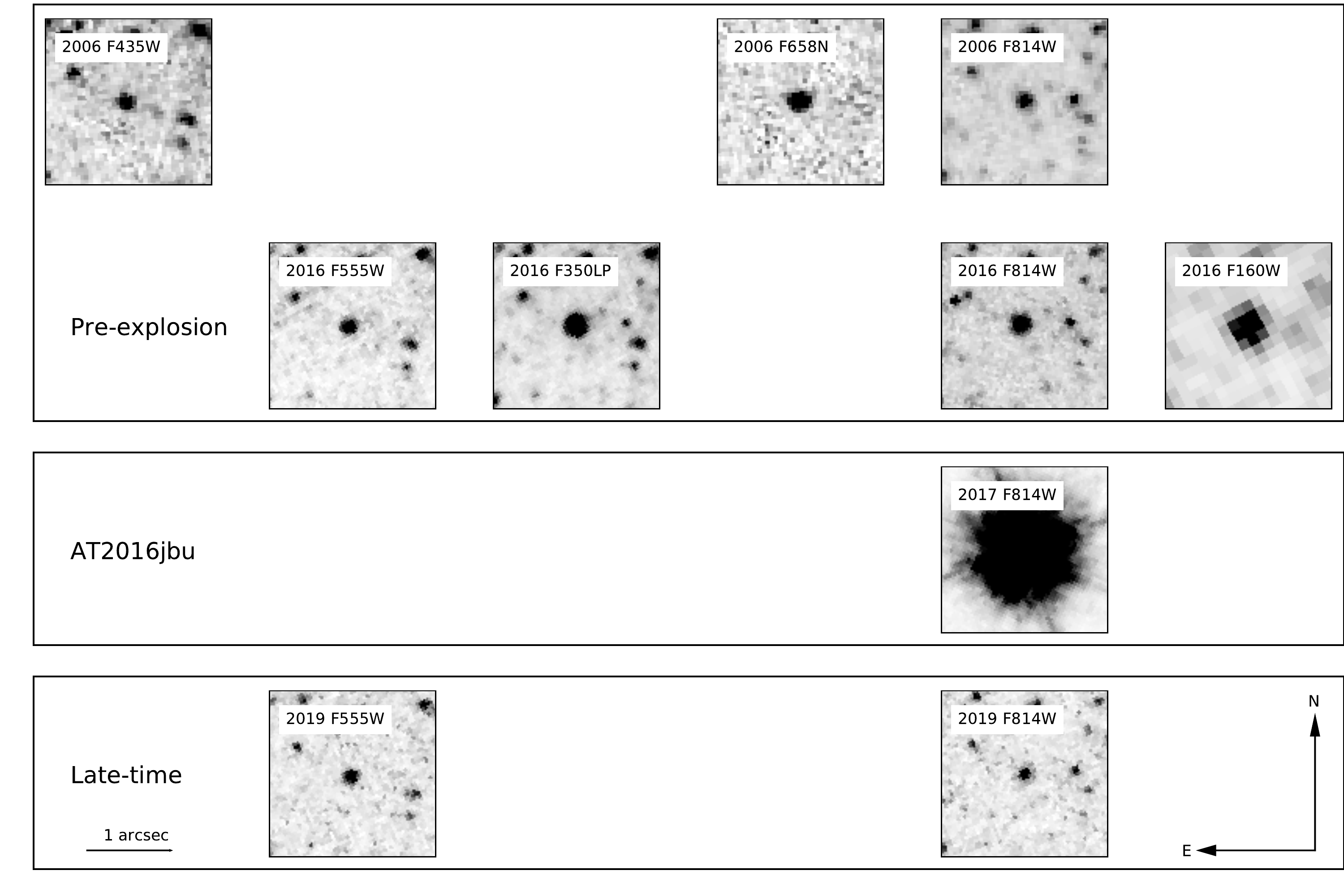}
\caption{2\arcsec$\times$2\arcsec\ cutouts of all HST images centered on the progenitor candidate for \jbu. Columns are ordered in wavelength from left to right.}
\label{fig:hst_cutouts}
\end{figure*}

In order to locate a progenitor candidate for \jbu, we aligned the {\it F814W}-filter image taken in 2017, when the transient was bright, to the ACS+{\it F814W} image from 2006, approximately ten years prior to discovery. Using 20 point sources common to both frames and within 20\arcsec\ of \jbu, we derive a transformation between the pixel coordinates with an root mean square (rms) scatter of only 12~milliarcseconds (pixel scale $\sim$~0.05~\arcsec/pixel). A bright source is clearly visible at the position, and we identify this as the progenitor candidate. The progenitor candidate is shown in Fig.~\ref{fig:hst_cutouts}, and is the same source as was identified by \citetalias{Kilpatrick2018}.

\begin{figure*}
\centering
\includegraphics[width= \textwidth]{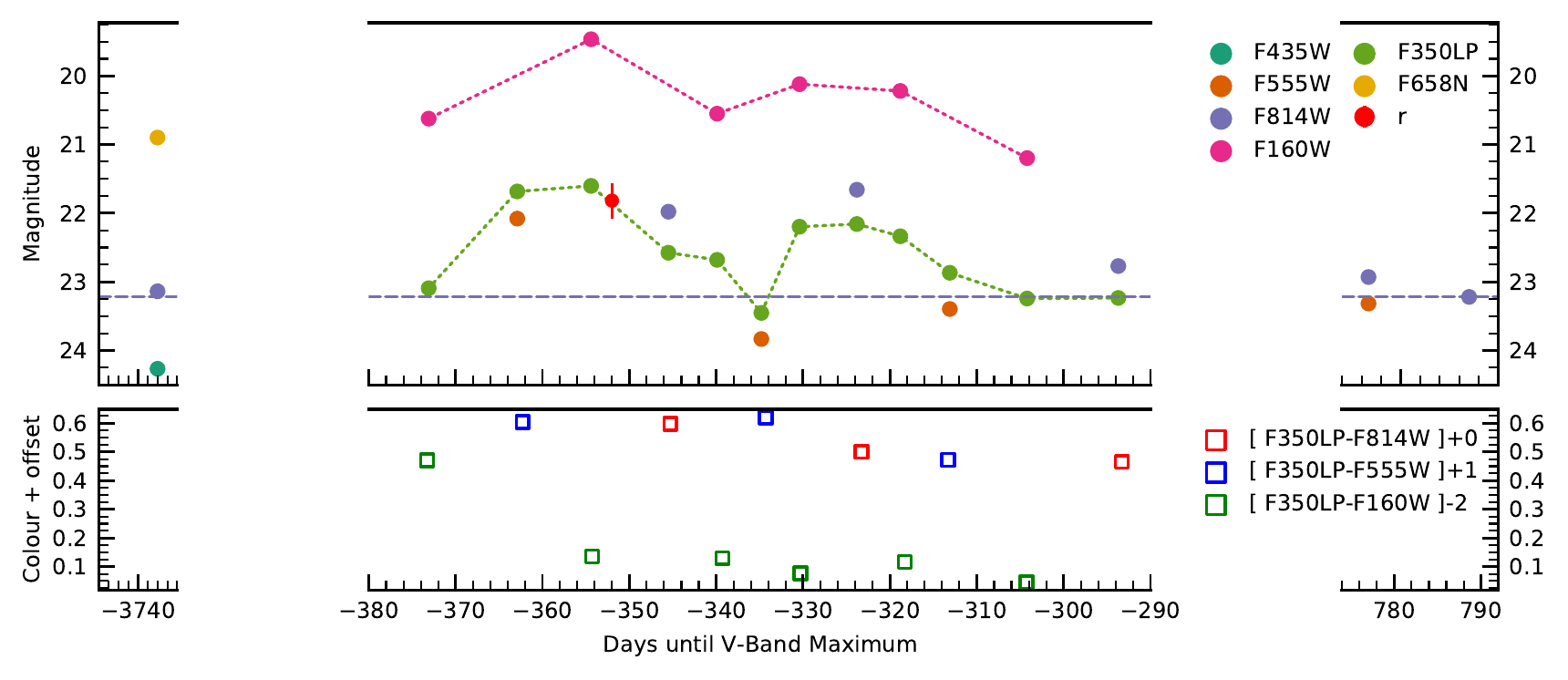}
\caption{Foreground extinction corrected {\it HST} lightcurves of \jbu and its progenitor are shown in the top panel. We also include a DECam \textit{r}-band detection at $-352$~d as a red filled circles with error bars. Error bars for {\it HST} measurements are smaller than the point sizes. The horizontal line is to guide the eye in comparing the late time ($\sim+2$ year) and pre-outburst ($\sim-10$~year) {\it F814W} magnitudes. We also plot the \textit{F350LP} and \textit{F160W} lightcurves with a line to help guide the eye. Colour curves, corrected for foreground reddening, are shown in the bottom panel. Colours as offset for legibility by the amounts stated in the legend.}
\label{fig:hst_lc}
\end{figure*}

We performed Point Spread Photometry (PSF) fitting on all {\it HST} images using the November 2019 release of the {\sc dolphot} package \citep{Dolp00}, with the instrument-specific ACS and WFC3 modules. In all cases, we performed photometry following the instrument-specific recommendations of the {\sc dolphot} handbook\footnote{\url{http://americano.dolphinsim.com/dolphot/dolphot.pdf}} regarding choice of aperture size. The WFC3 images were taken at two distinct pointings, and each set were analysed separately, otherwise each contiguous set of imaging with a particular instrument were photometered together, using a single deep drizzled image as a reference frame for source detection. Examination of the residual images after fitting and subtracting a PSF to sources in the field revealed no systematic residuals, indicating satisfactory fits in all cases. We show the HST photometry for \jbu in Fig.~\ref{fig:hst_lc}.

We find that the photometry reported by \citetalias{Kilpatrick2018} is fainter than what we measure, with a difference of $\sim0.5$~mag in \textit{F350LP}. We compared our measured \textit{F350LP} magnitudes and those of \citetalias{Kilpatrick2018} to the values reported in the Hubble Source Catalog (HSC; \citealp{Whit16}). As the magnitudes reported in the HSC are in the AB~mag system, we applied the conversion from AB to Vega mag before comparing to our photometry. The HSC \textit{F350LP} magnitudes are consistent with what we report here, and we also see the same variability for the progenitor candidate. The cause of the difference between our photometry and that of \citetalias{Kilpatrick2018} hence remains unknown.

\begin{figure}
\includegraphics[width= \columnwidth]{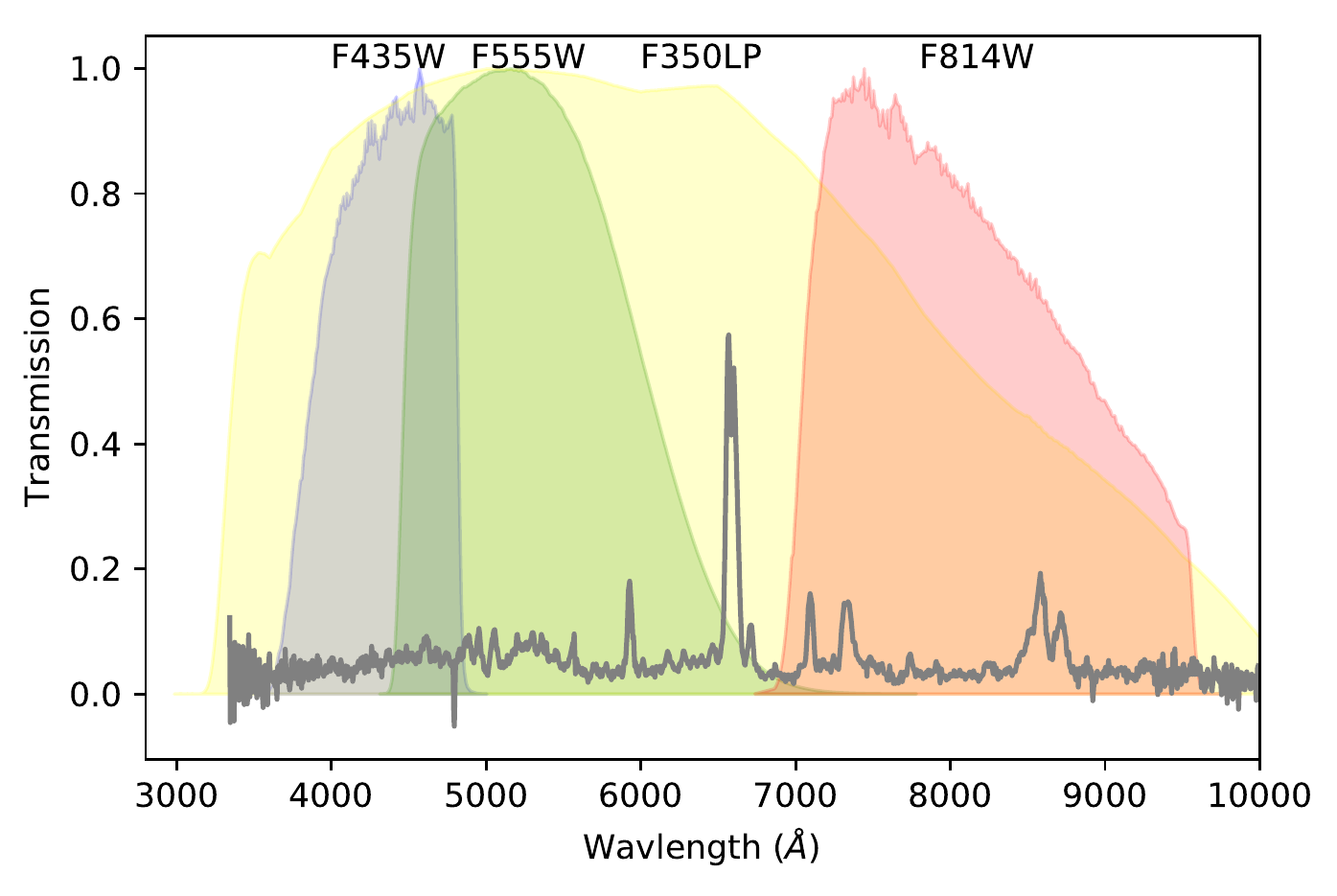}
\caption{{\it HST} filters used for pre-explosion lightcurve, compared to the $+271$~d spectrum. Only {\it F350LP} covers the strong H$\alpha$ emission seen at this epoch. }
\label{fig:hst_filters}
\end{figure}

We note that the broad-band photometry from {\it HST} is more than likely affected by the strong emission in H$\alpha$. In Fig.~\ref{fig:hst_filters} we show the throughput of the {\it HST} filters compared to a late-phase spectrum of \jbu. The long-pass \textit{F350LP} filter will contain flux from H$\alpha$. Fortuitously, H$\alpha$ falls in the low-throughput red wing of the \textit{F555W} filter, where it will have negligible effect. To verify this, we used {\sc synphot} \citep{pey_lian_lim_2020} to perform synthetic \textit{F555W}-filter photometry on the $+271$~d spectrum of \jbu, and on the same spectrum where H$\alpha$ has been excised. The latter returns a magnitude that is only 0.05~mag fainter than the former, and so the \textit{F555W} filter is not significantly affected by line emission. 

The progenitor is relatively red, bright, and shows significant variability over timescales of $\sim$weeks. Correcting for foreground extinction, in 2006 the progenitor candidate had an absolute magnitude in \textit{F814W} = $-8.46\pm0.06$ and an $F435W-F814W$ colour of 1.13$\pm$0.04~mag. This colour is consistent with a yellow hyper-giant (YHG), and corresponds to a blackbody temperature of 6500~K \citep{DrillingLandolt}. 

However, the narrow-band $F658N$ magnitude, which covers H$\alpha$, is much brighter than would be expected. This indicates that even ten years before the eruption or explosion of \jbu its progenitor was characterised by strong H$\alpha$ emission.

In early 2016, between seven and ten months prior to the start of \eventA, NGC~2442 was observed repeatedly with WFC3 in \textit{F350LP}, \textit{F555W}, \textit{F814W} and \textit{F160W}. This dataset gives us a unique insight into the variability of the quiescent progenitor prior to explosion. We see that even in quiescence (arbitrarily defined as when the progenitor is fainter than mag$\sim$-10), the progenitor displays strong variability. In particular in the best-sampled \textit{F350LP} lightcurve, the progenitor varies in brightness by 1.9~mag in only 20~days. As discussed by \citetalias{Kilpatrick2018}, such rapid variability is hard to explain (although there is some similarity to the fast variability seen in the pre-explosion lightcurve of \ip; \citealp{Pastorello2013}). While it is impossible to know if the variability is periodic on the basis of the short time coverage available for \jbu, if it is periodic then the apparent period is around 45~days (found via a low order polynomial fit to the \textit{F350LP} lightcurve).

The variability seen in \textit{F350LP} in early 2016 is also seen in other bands, which appear to track the same overall pattern of brightening and fading. Fig.~\ref{fig:hst_lc} shows the colour evolution of \textit{F350LP-F555W}, \textit{F350LP-F814W}, \textit{F350LP-F160W}. In all cases (with the exception of the earliest \textit{F350LP-F160W} colour, which is likely due to a spurious \textit{F350LP} magnitude) we see a relatively minor colour change over three months. In fact, it is possible that the apparent small shift towards bluer colours is simply due to H$\alpha$ growing stronger, which would cause the \textit{F350LP} magnitude to appear brighter, rather than any change in the continuum temperature.

At late times the progenitor candidate for \jbu is still present. In 2019, over two years since the epoch of maximum light, a source is found at approximately the same \textit{F814W} magnitude as was seen in 2006. It is unlikely that this source is a compact cluster, as the pre-explosion photometric variability can only be explained if a single star is contributing most of the flux. Moreover, we compared the 2006 \textit{F814W} and 2019 \textit{F814W} images, and find that the position of the source is consistent to within 17~mas between the two epochs. This implies that the same source is likely dominating the emission at both epochs, and if there is an underlying cluster it must be much fainter than the progenitor source.

\subsection{Physical properties of the progenitor}\label{sec:physical}

\begin{figure}
\includegraphics[width=1\columnwidth]{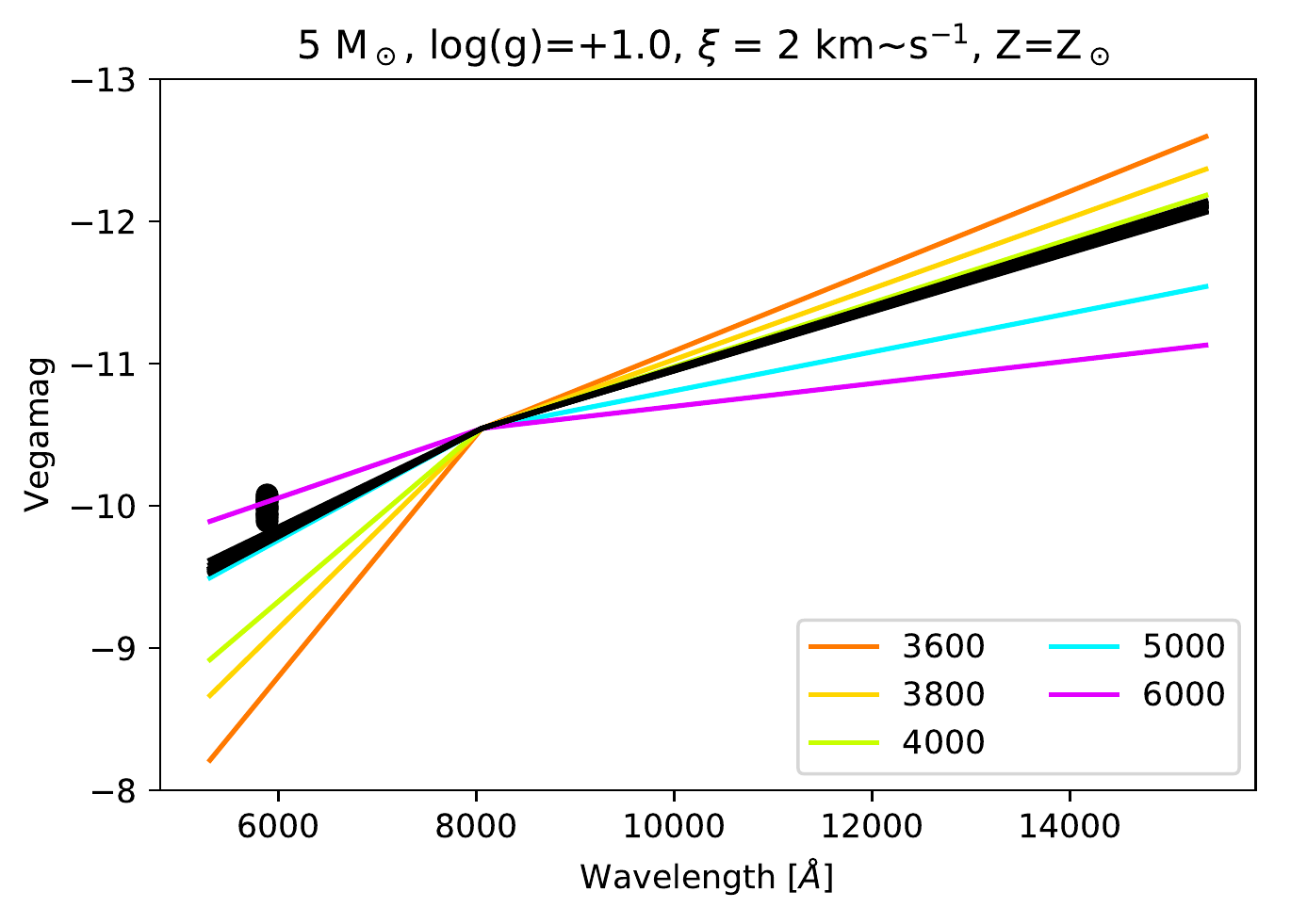}
\includegraphics[width=1\columnwidth]{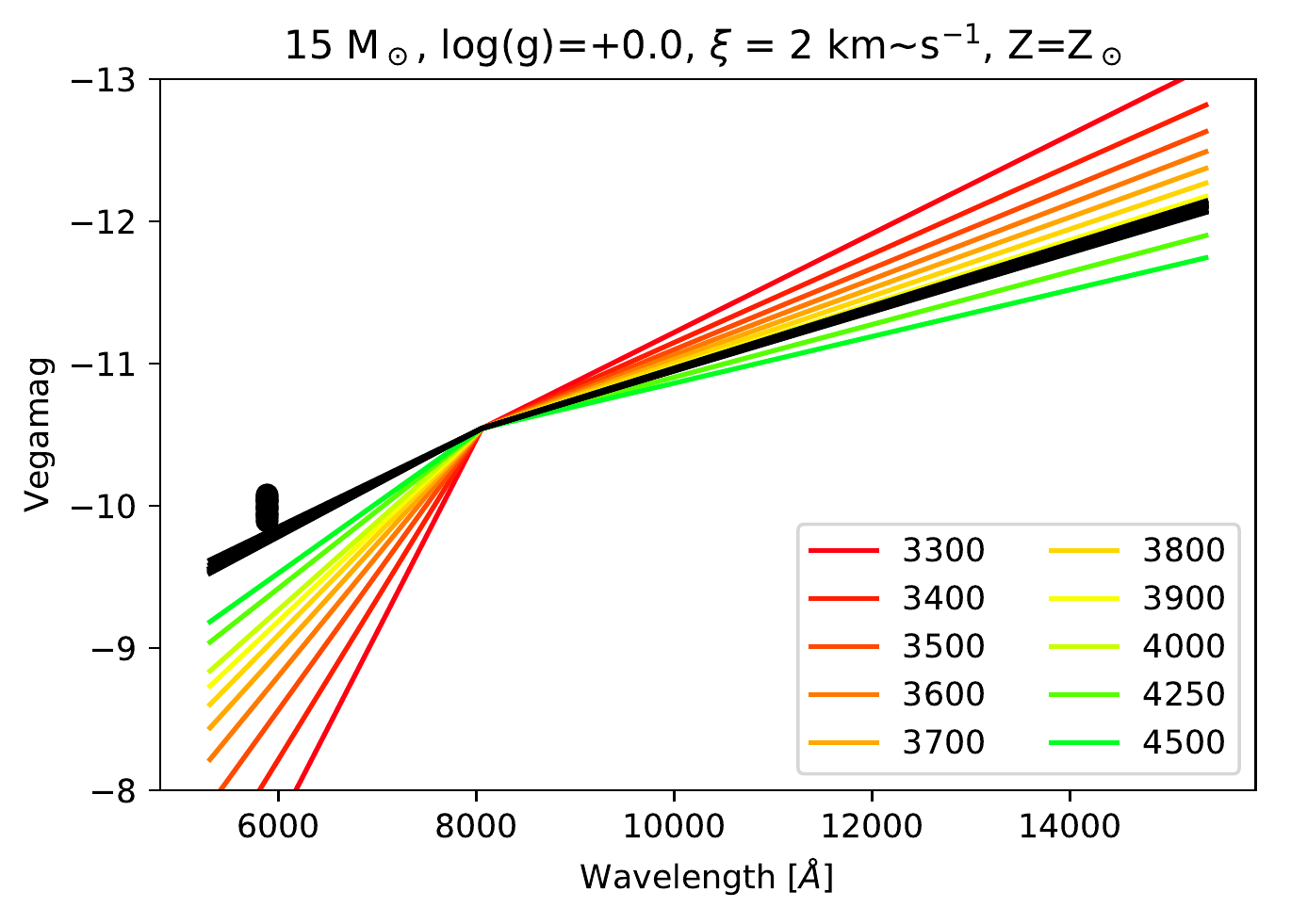}
\caption{{\it HST} SEDs for \jbu based on the early 2016 WFC3 imaging are shown in black. All SEDs have been shifted so that their \textit{F814W} magnitudes match, as discussed in the text. The \textit{F350LP} filter magnitudes have not been included in the SED as they are strongly affected by H$\alpha$ emission. We also plot a number of SEDs derived from \textsc{MARCS} models. In the lower panel we show the 15~\msun\ MARCS models appropriate to cool red supergiants. As this model grid does not extend above 4500~K, we also plot a set of 5~\msun\ models with slightly higher log(g) in the upper panel. All models have been shifted so that they match the \textit{F814W} filter magnitude of the progenitor, and we can see that while the cooler models can match the NIR part of the SED, hotter temperatures are required to match the optical.}
\label{fig:hst_SED}
\end{figure}

In order to determine the luminosity and effective temperature of the progenitor of \jbu, we consider the WFC3 photometry taken in early 2016. As a first step, we normalize out the variability seen over this period so that we can build an SED from photometry taken in different filters at different epochs. To do this, we fit a linear function to the colour curves of our {\it HST} observations. We disregard the first epoch for the {\it F350LP}$-${\it F160W} colour (which is significantly redder than the other epochs); this measurement is unreliable as the progenitor was affected by bad pixels in two of the three individual exposures. We then use the fitted functions to interpolate or extrapolate the magnitude of \jbu in \textit{F555W}, \textit{F814W} or \textit{F160W} as necessary. Finally, we shift the SEDs up or down in magnitude so that they all have the same \textit{F814W} magnitude as the 2006 value. The resulting normalised progenitor SEDs can be seen in Fig.~\ref{fig:hst_SED}.

In order to determine a progenitor temperature from the observed SED, we compare to MARCS stellar atmosphere models \citep{Gust08}. We used the {\sc pysynphot} package to perform synthetic photometry on the surface fluxes of the models and hence calculate their magnitude in each of the \textit{F555W}, \textit{F814W}, and \textit{F160W} filters. We shifted each model so that it matches the 2006 MW extinction corrected \textit{F814W} absolute magnitude of the progenitor. In the lower panel of Fig.~\ref{fig:hst_SED} we compare to the spherically-symmetric MARCS models for 15\msun\ red supergiants (RSGs; log(g)=0) at solar metallicity. While we can see that the models provide a reasonable agreement, it is clear that the warmest model (at 4500~K) is still too red to match the \textit{F555-F814W} colour of the progenitor, implying that the progenitor is hotter than this. Conversely, the 4000~K model provides a good match to the \textit{F814W-F160W} colours of the progenitor. As the 15~\msun\ super-giant models cover a relatively small temperature range, we also explored the 5~\msun\ spherically symmetric MARCS models at log(g)=1.0 which span a broader range (upper panel in Fig.~\ref{fig:hst_SED}). We find that a 5000~K model can reproduce the optical colours of the progenitor, while the NIR is better matched with a cooler 4000~K model. 

While \jbu does not appear to suffer from high levels of circumstellar extinction around maximum light, we cannot exclude the possibility that the progenitor colours are caused by close-in CSM dust that was subsequently destroyed. To explore this possibility, we used the {\sc dusty} \citep{Ivez97} code to calculate observed SEDs for a grid of progenitor models allowing for different levels of CSM dust. {\sc dusty} solves for radiation transport within a dusty medium.

\begin{figure*}
\centering
\includegraphics[width=\textwidth]{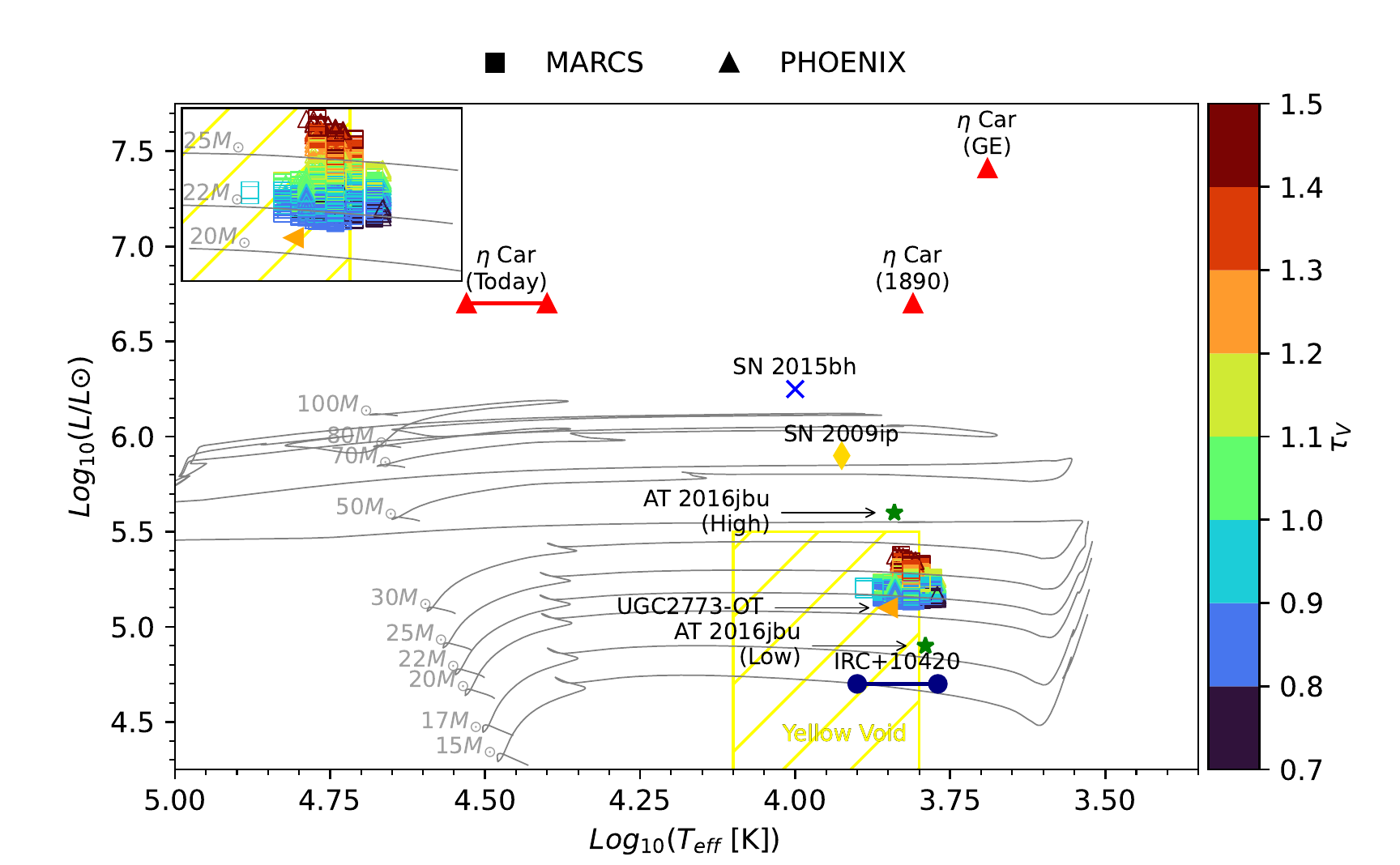}
\caption{ Hertzsprung–Russell (HR) diagram showing single star evolutionary tracks from \textsc{BPASS} \citep{Eldridge2017,Stanway2018}. We include \ip at log$(L/L_{\odot})=5.9$ and log$(T_{eff})=3.92$ \citep{smith10,Foley2011}, as well as \bh \citep{Boian2018}, \irc \citep{Klochkova16}, and UGC2773-OT \citep{smith2015}. \etacar is plotted (red triangles) at several phases given in parentheses \citep{Rest_2012}. We include the progenitor estimates for \jbu from \citetalias{Kilpatrick2018}, in both the ``low'' and ``high'' states as green stars . We highlight the \textit{Yellow-Void} between 7000~K and 10000~K \citep{deJager1998} and include the output of our {\sc dusty} modelling for \jbu using \textit{PHOENIX} models (multi-coloured triangles) and \textit{MARCS} models (multi-coloured squares). The colour of each point corresponds to its optical depth ($\tau_\nu$) which is provided on the colour bar on the right. We include an inset of the region around the progenitor in the top left of the plot.}
\label{fig:HR_diagram}
\end{figure*}

Since a dust-enshrouded progenitor could be hotter than the range of temperatures covered by the \textsc{MARCS} model grid, we used the \textsc{PHOENIX} models\footnote{\url{http://phoenix.astro.physik.uni-goettingen.de}} \citep{Husser_2013} as our input spectra. The \textsc{PHOENIX} models cover the temperature range from 6000--12000~K in 200~K increments, and have log(g) between 1 and 2 dex. \textsc{MARCS} models covering a temperature range from 2600--7000~K in 100~K increments and log(g) between 1 and 2 dex were also tested as input to {\sc dusty}. These models were then processed by {\sc dusty}, assuming spherically symmetric dust comprised of 50 per cent silicates and 50 per cent amorphous carbon. The dust density followed a $r^{-2}$ distribution, with a radial extent varying between 1.5 and 20 times the inner radius of the dust shell. The dust mass is parameterized in terms of the optical depth in $V$-band, $\tau_V$, which varied between 0 and 5. We expect the dust temperature to be relatively hot \citep{Foley2011,smith2013b}. We vary the dust temperature at the inner dust boundary between 1250-2250 K. For each temperature and dust combination, we calculated synthetic $F555W-F814W$ and $F814W-F160W$ colours, and compared to the foreground extinction corrected colours of the \jbu progenitor. In Fig.~\ref{fig:HR_diagram} we plot all models that have colours within 0.1~mag of the progenitor. 

% We find that we are able to match the progenitor colours with models with temperatures of between $10^{3.8}$ and $10^{3.9}$~K, for a circumstellar dust shell with optical depth $\tau_V$ between 1.6 and 2.6.

We find that we are able to match the progenitor colours with models with temperatures of between $10^{3.7}$ and $10^{3.9}$~K, for a circumstellar dust shell with optical depth $\tau_V$ between 0.7 and 1.5, and a dust temperature between 1500-2000K, in agreement what was seen in the environment of \ip \citep{smith2013b}, as well as the SN Imposter, UGC2773-OT \citep{Foley2011}. Additionally we find little influence of the radial extent of the dust on matching models. 

We calculated a luminosity for each of these models by integrating over its spectrum, and find that the progenitor had a luminosity log$(L)$ between 5.1 and 5.3~dex (depending on temperature and extinction). Comparing to the {\sc BPASS} single star evolutionary tracks at Solar metallicity in Fig.~\ref{fig:HR_diagram}, we find that these correspond to approximately the luminosity of a 22--25~\msun star as it crosses the HR diagram to become a RSG. We plot the {\sc DUSTY} models that match our progenitor measurements in Fig.\ref{fig:HR_diagram}.

\section{Evidence for Dust}\label{sec:dust}
\begin{figure}
\centering
\includegraphics[width=\columnwidth]{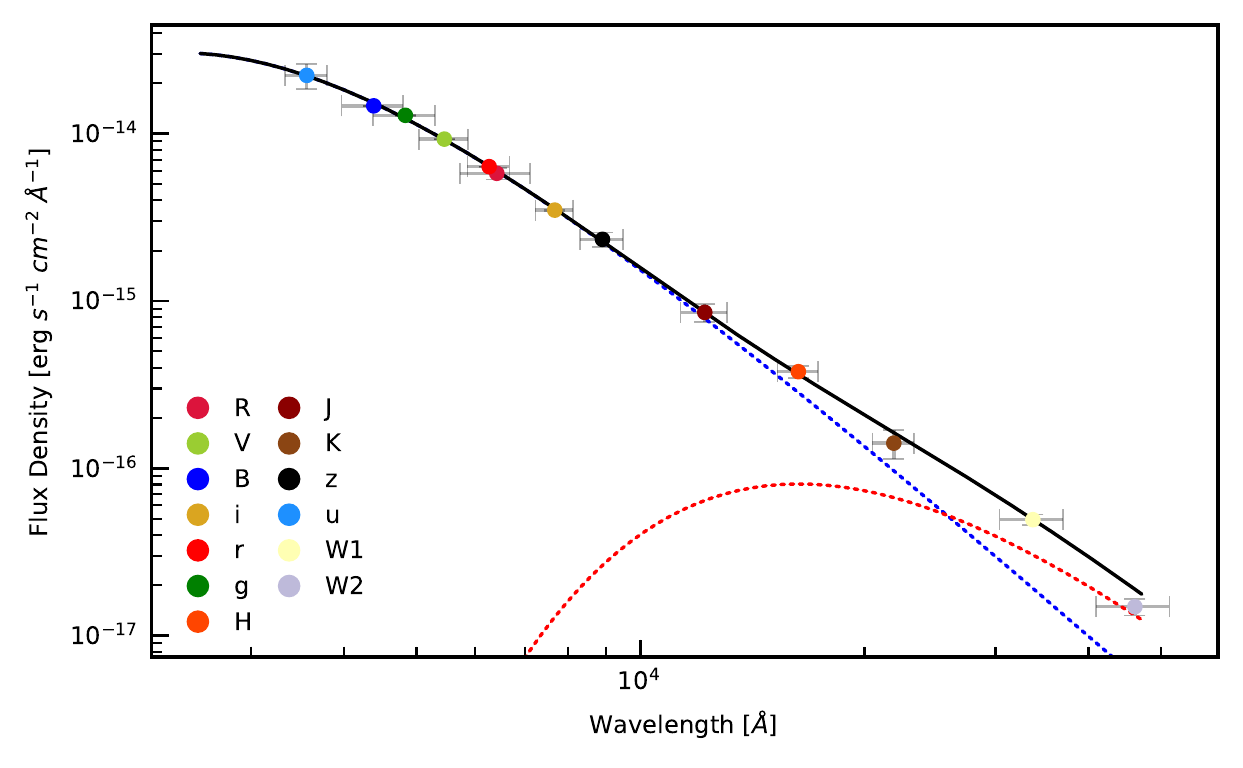}
\caption{SED fit of \jbu at -11~days before \eventB maximum. Extinction corrected photometry are grouped to 1 day bins and weighted averaged. Flux errors are given as standard deviation of bins. Horizontal error bars represent approximate filter band-pass. Hot blackbody is given in blue, the cooler blackbody in red, and the black line is compound model. We note a similarity the SED for \ip presented in \citet{Margutti2014}.}
\label{fig:sed_fit}
\end{figure}
We present a SED model fitted to our $-11$ day dataset in Fig.~\ref{fig:sed_fit}. We fit at this phase as it has the broadest wavelength coverage without the need for interpolation. We fit two blackbody models to the photometric points; one representing a hot photosphere, and the second fitted to the IR excess seen in $H$, $K$, $W1$, and $W2$. A single blackbody does not fit observations seen at $-11$~d before maximum. Allowing for a second cooler blackbody at a larger radius gives a model that fits the data well. 
This additional blackbody is consistent with warm dusty material at a distance of 170~AU and a temperature of $T_{\rm BB}\sim$1700~K. This material provides an additional luminosity of \SI{2.7e7}{L_\odot}. The hot blackbody has a radius of 36~AU, a temperature of $T_{\rm BB}\sim$~12000~K and has an integrated luminosity of \SI{1.3e8}{L_\odot} and represents $R_{\rm BB}$ at this time. We find a dust mass of $M_{dust}\approx2.27\times10^{-6}$~\msun\ \citep[Using eq.1 from][]{Foley2011}. In comparison \citet{smith13} finds a lower dust mass of $(3-6) \times10^{-7}$~\msun\ for \ip. Additionally we note a similarity to the SED for \ip presented in \citet{Margutti2014}. The IR excess may be caused by thermal radiation of pre-existing dust in the CSM re-heated by an eruption at the beginning of \eventB, i.e. an IR echo.
We can compute the radius within which any dust will be evaporated/vaporised at the phase of our SED fitting. The radius of this dust-free cavity is given by:
\begin{equation}\label{eq:cavity_radius}
R_c = \sqrt{\frac{L_{SN}}{16~\pi~\sigma~T_{evap}^4~\langle Q \rangle}}
\end{equation}
\noindent where $R_c$ is the cavity radius, $L_{SN}$ is the luminosity of the transient, taken to be \SI{1.3e8}{L_\odot}, $\sigma$ is the Stefan-Boltzmann constant and $\langle~Q~\rangle$ is the averaged value of the dust emissivity. Assuming radiation is absorbed with efficiency $\sim$unity by the dust, we find a cavity radius of $\sim$~245~AU for graphite grains ($T_{evap} = 1900~K$) and $\sim$~400~AU for silicate grains ($T_{evap} = 1500~K$). Both values are significantly larger than what we find from our warm blackbody radius ($\sim$~170~AU).
A dust destruction radius larger than the blackbody radius of our putative warm dust component appears at first glance to be inconsistent. To ameliorate this we suggest that the dust may not be homogeneously distributed, and could be in either optically clumps or an aspherical region that provides some shielding from evaporation. Over time, we expect that the dust is further heated and destroyed during the rise to \eventB maximum. We find that by maximum brightness that this additional blackbody component is no longer needed, suggesting that the dust causing this NIR excess has been destroyed.
As discussed in \paperI, as well as \citetalias{Kilpatrick2018}, there are \textit{Spitzer}+IRAC observations of the progenitor site of \jbu, which show tentative detections in 2003 and 2018. Using 2003 Spitzer/IRAC and 2016 {\it HST}/F160W observations, \citetalias{Kilpatrick2018} find fits consistent with a compact dusty CSM component with mass $M_{dust} \approx \SI{7.7e-7}{M_{\odot}}$ at 72~AU. This may represent a dusty shell that is later seen as our 170~AU warm blackbody. However, due to the time frame between Spitzer/{\it HST} observations there are large uncertainties on dust parameters from \citetalias{Kilpatrick2018}. Fitting Spitzer data only gives a slightly higher $M_{dust}$ value of $\sim$~$10^{-6}$~\msun\ at 120~AU. Due to the erratic variability seen in \jbu it is uncertain as to whether these dust shells are the same, as \jbu may have a stratified CSM environment resulting from successive outbursts. 
Although there is strong evidence for pre-existing dust, we do not see any signature for newly formed dust in the environment around \jbu \citep{Meikle_2007,smith2008,Smith_2011}. We see no NIR excess in late time $J$ and $K$ bands in late time photometry nor an IR excess evident in spectra. Furthermore, there is no blue-shift in the core emission component in H$\alpha$ (\paperI), which is another indicator of newly formed dust. 

%%%%%%%%%%%%%%%%%%%%%%%%%%%%%%%
%%%%%%%%% Environment %%%%%%%%%
%%%%%%%%%%%%%%%%%%%%%%%%%%%%%%%

\section{The environment of \jbu}\label{sec:enviroment}

Along with direct detections of progenitors, analysis of the resolved stellar population in the vicinity of a SN has also been used to infer the progenitor age and hence initial mass \citep{Goga09,Maun17,Will18}. An advantage to this technique is that it will not be affected by any peculiar evolutionary history or variability of the progenitor that may cause it to appear less or more massive than it truly is. On the other hand, using the environment around a SN is an indirect proxy for the progenitor age, and is predicated on the assumption that the local stellar population is coeval. This method is also complicated by possible contamination from other stellar populations from multiple star formation episodes.

\subsection{Hubble Space Telescope imaging of the environment}

In order to study the population in the vicinity of \jbu we require sources to be matched between different filter images. While this is straightforward for bright sources such as the progenitor of \jbu it is more challenging for fainter or blended sources, especially when images have different pixel scales or orientations. We hence re-ran the photometry on a subset of the {\it HST} images (\textit{F435W}, \textit{F658N} and \textit{F814W} from 2006 Oct 20; \textit{F350LP} and \textit{F555W} from 2016 Jan 31), using a single drizzled ACS \textit{F814W} image as the reference image for all filters.

We chose a projected radius of 150~pc (1.48\arcsec) around \jbu as a compromise between identifying sufficient stars to be able to constrain the population age and ensuring we are still sampling a local population that is plausibly coeval with the progenitor. We also create a less restrictive catalog of sources within a projected distance of 300~pc from \jbu, as well as a more limited catalog of sources within 50~pc. After applying cuts to select only sources with a point source PSF, {\sc dolphot} detects 84 sources at S/N$>3$ within 150~pc of \jbu, and 255 sources within 300~pc.

\begin{figure*}
\centering
\includegraphics[width=\textwidth]{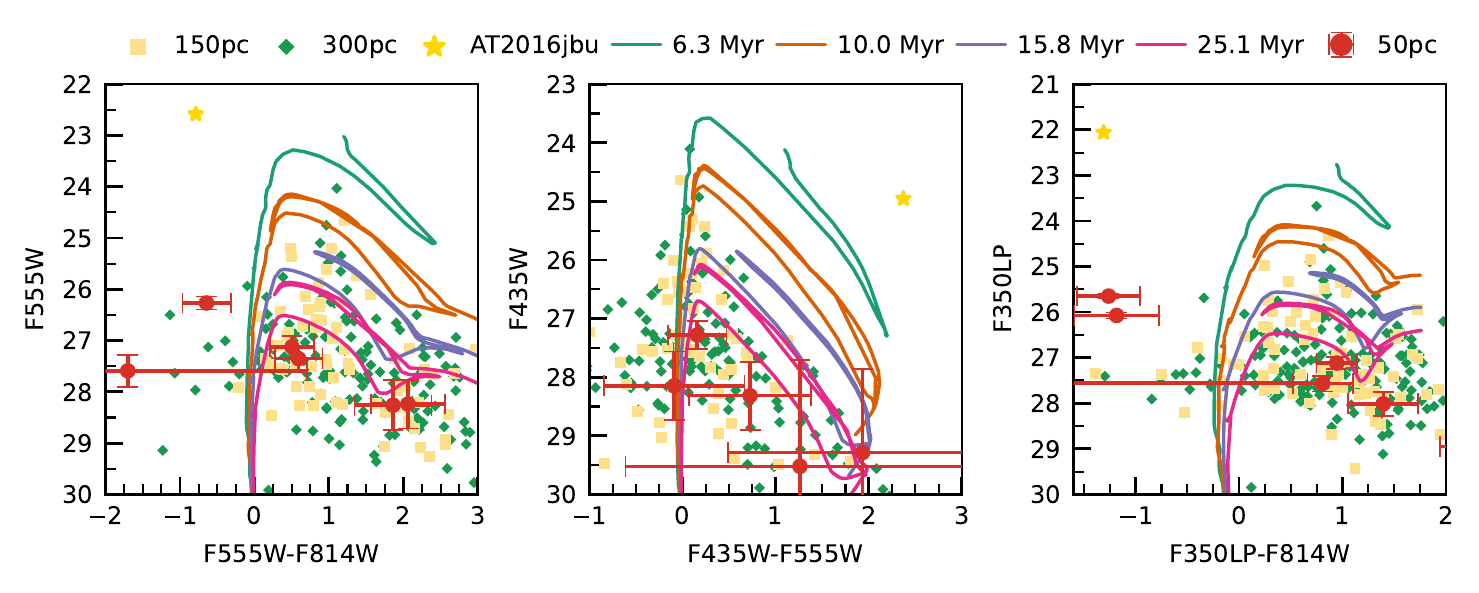}
\caption{Color magnitude diagram (CMD) of the stellar population around the site of \jbu. We show three different colour combinations each with {\sc PARSEC} isochrones with population ages (solid colored lines) given in the upper legend. Yellow squares are point sources within 150~pc and green diamonds are within 300~pc, while sources within 50~pc of the progenitor are plotted in red with error bars. The progenitor of \jbu from the early 2016 {\it HST} observations is given as a gold star in each panel.}
\label{fig:hst_isochrones}
\end{figure*}

\begin{figure}
\includegraphics[width=\columnwidth]{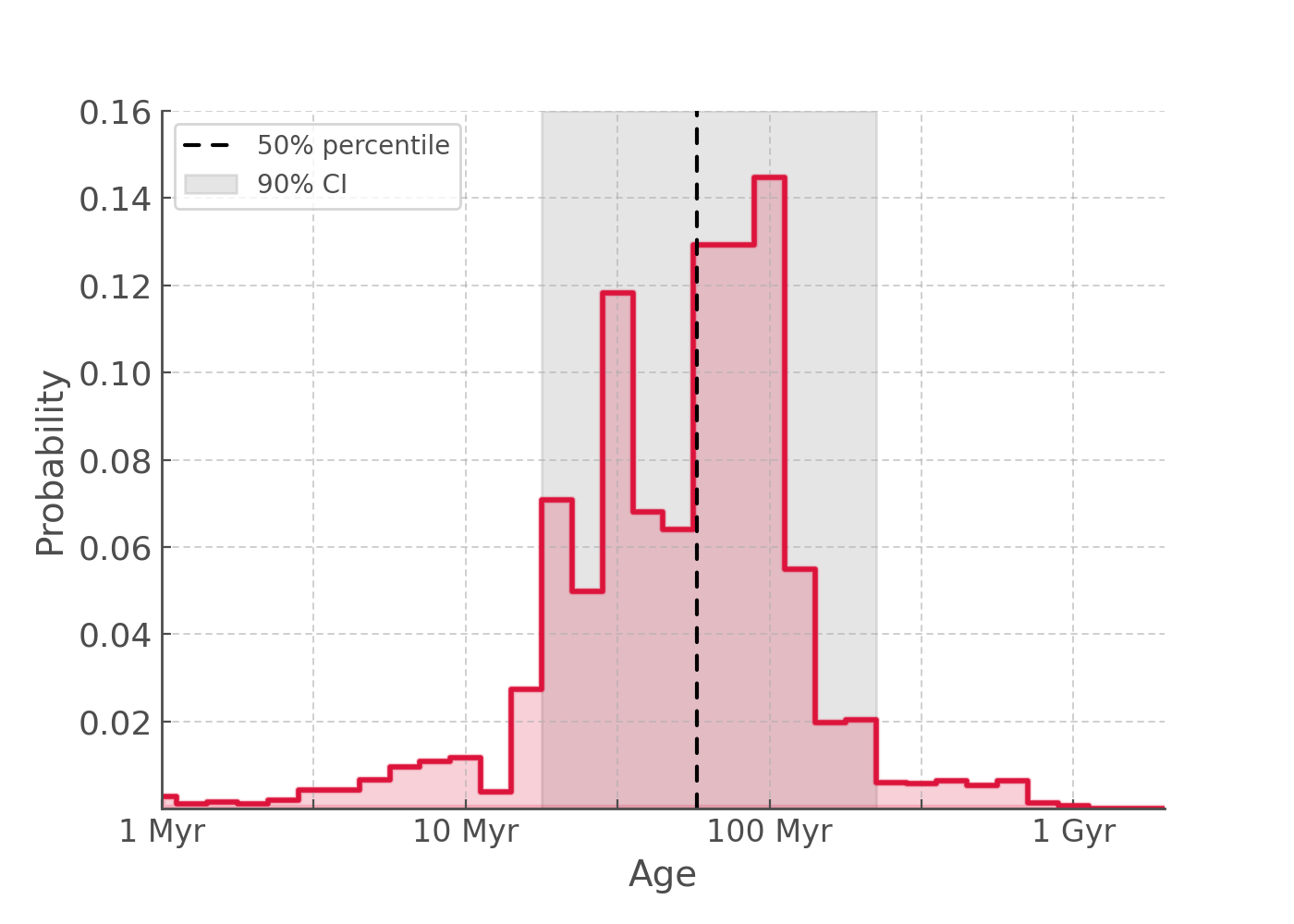}
\caption{Probability distribution of the age of the 150~pc stellar neighbourhood of \jbu found using {\sc AgeWizard}. The 90 percent confidence interval is highlighted in grey.}
\label{fig:age_wizard}
\end{figure}

In Fig.~\ref{fig:hst_isochrones} we compare our 50~pc, 150~pc and 300~pc populations to a set of {\sc PARSEC} isochrones\footnote{\url{http://stev.oapd.inaf.it/cgi-bin/cmd}} \citep{Marigo2017,Bressan12} in three different filter/colour combinations. We use the most recent version of the {\sc PARSEC} models \citep[version 1.2S;][]{Chen2015}, and for the purposes of the comparison we have applied our foreground reddening and distance modulus to the {\sc PARSEC} models.

The progenitor of \jbu clearly stands out from the local population, both in terms of its bright apparent magnitude and unusual colours. The colour of \jbu should not be compared to these isochrones; not only will the $F350LP$ filter be strongly affected by H$\alpha$ emission, but as the various filter combinations plotted do not come from contemporaneous data, the variability seen in the progenitor will significantly affect the apparent colour.

Turning to the 150~pc population, it is clear that no source is found to be brighter than the 10~Myrs isochrone, constraining the population to be older than this age. We find a similar result looking at the wider environment within 300~pc of \jbu as well as the closer in population within 50~pc.

Using {\sc AgeWizard} and {\sc BPASS} models \citep{Eldridge2017,stevance2020, stevance2020b}, we obtain a probability distribution for the age of the resolved stellar population within 150~pc around \jbu (see Fig. \ref{fig:age_wizard}). The 90 percent confidence interval is found to be $15$--$200$~Myrs. Additionally, we can ascertain that the neighbouring population of \jbu is older than 10~Myrs (5~Myrs) with over 95 (99.8) percent confidence.

Therefore, there is no evidence for a very young environment which would be expected for a 80 (or even 150)~\msun\ progenitor as proposed for \ip and \etacar \citep{smith10,Foley2011}.

\subsection{\textit{MUSE}-ing on the local environment}\label{sec:muse}

\begin{figure*}
\centering 
\includegraphics[width=\textwidth]{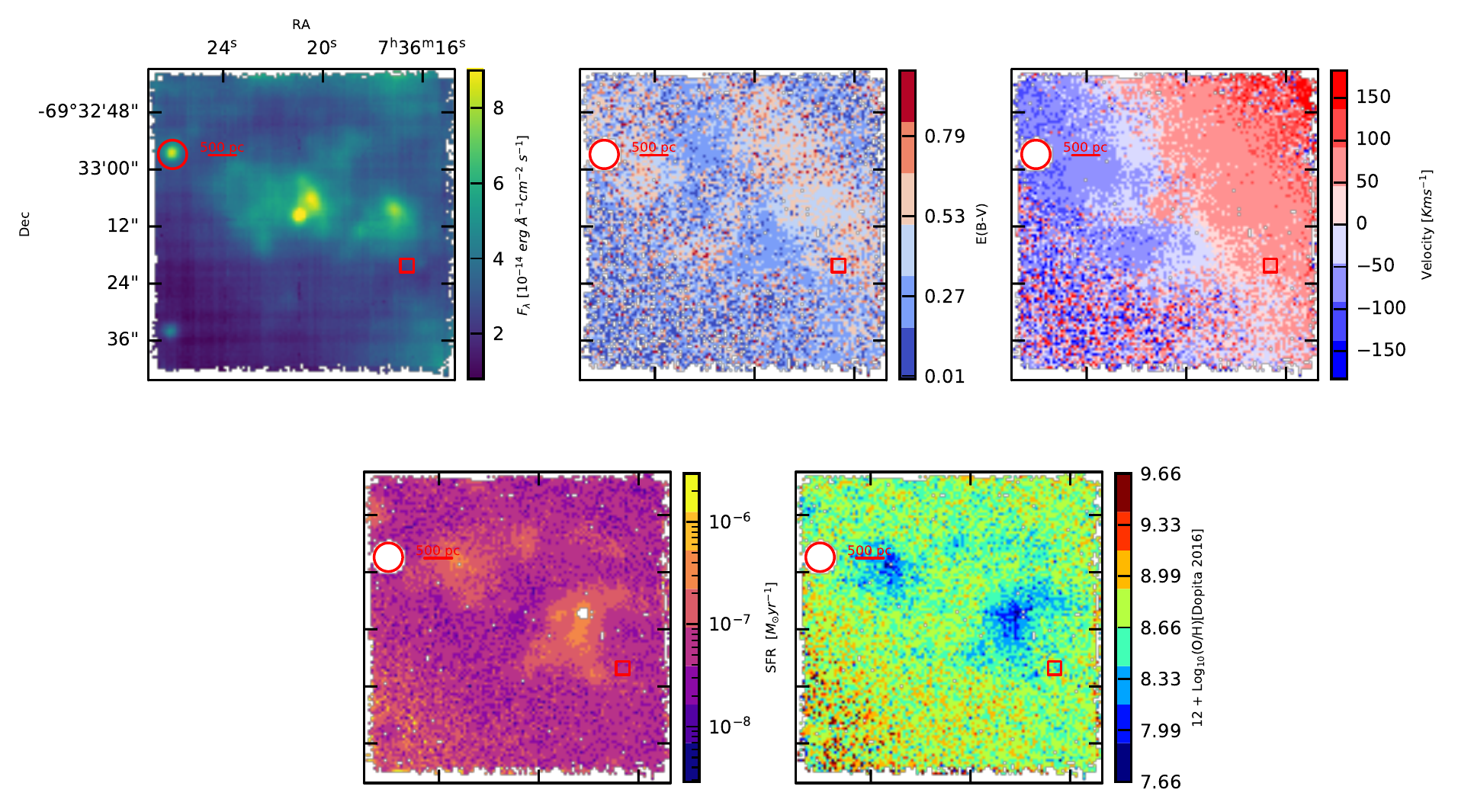}
\caption{IFU analysis of the environment of \jbu. The spectral cube was corrected for Milky Way extinction and redshift. Observations were taken on 2017 Dec 2 ($+303$~d). Data is orientated such that north is up and east is to the left. Included in each panel is a horizontal scale bar showing 500~pc. We include a white light image (5000--7000\AA) (top left), an extinction map (top middle) based on \citet{Dominguez2013}, a velocity field plot from H$\alpha$ corrected for recessional velocity (top right), star formation rate based on \citet{Kennicutt1998} (bottom left) and a metallicity map (bottom right) based on \citet{Dopita2016}. The location of \jbu is marked with a red circle of radius 3~\arcsec. We also include the location of SN~1999ga as a square to the south west of \jbu. Data is not shown where EW $<$ 1~\AA\ or within 3~\arcsec\ of \jbu.}
\label{fig:muse_enviroment}
\end{figure*}

We further investigate the nature of \jbu by looking at its local environment in Integral Field Unit (IFU) data. \jbu was observed on 2017 Dec 2 ($+303$~d) using the VLT equipped with the MUSE instrument in Wide Field Mode. The date cube was obtained as part of a survey of SN late-time spectra in conjunction with the AMUSING survey of SN environments \citep{Galbany2016,Kuncarayakti2020}. We downloaded the pre-calibrated data cubes from the ESO archive and present our data analysis for the environment around \jbu in Fig.~\ref{fig:muse_enviroment}.

\begin{figure*}
\centering
\includegraphics[width=\textwidth]{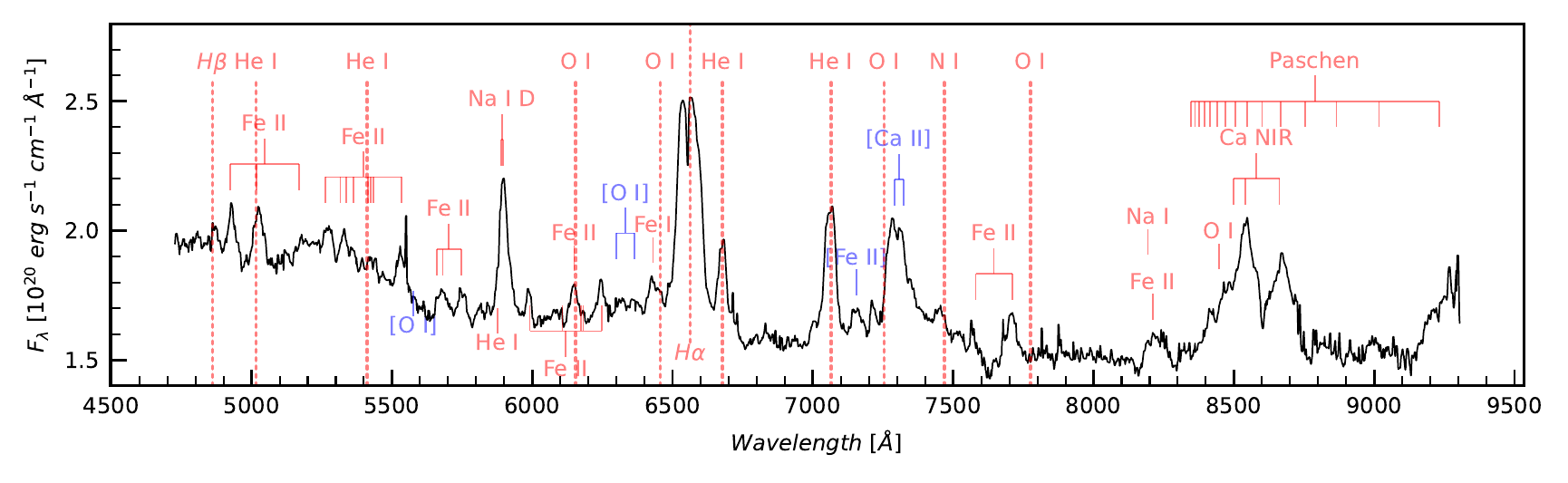}
\caption{Extracted spectrum of \jbu from VLT+MUSE. Spectrum was extracted using a 1\arcsec\ aperture at the transient position and corrected for redshift and Galactic extinction. We marks strong emission features in red and several forbidden transition lines are marked in blue. The transient appears relatively blue even at $\sim$~+10 months, a possible sign of ongoing interaction.}
\label{fig:muse_spectra}
\end{figure*}

We fit for spectral features at each spaxel using a Gaussian emission profile with a linear pseudo continuum over a small wavelength range. For measuring the ratio of H$\alpha$ and H$\beta$ for the extinction map we constrain the ratio of the two emission lines such that $H{\alpha} / H{\beta} ~\le~2.85$ (Case B recombination). To exclude the effects of \jbu on the analysis, we exclude any pixel within 3~\arcsec\ of \jbu. We do not account for any stellar absorption effects and as such, values here are lower limits. For completeness we include the extracted spectrum of \jbu in Fig.~\ref{fig:muse_spectra}.

We show the extinction map across the field of view (FOV) using the method in \citet{Dominguez2013}, measured using the Balmer decrement. A proxy for the star formation rate (SFR) is measured using L$_{H\alpha}$ \citep{Kennicutt1998}. L$_{H\alpha}$ was corrected for extinction using the Balmer decrement \citep{asari2020sum}. We also plot a metallicity map using the metallicity indicators given by \citet{Dopita2016}.

Fig.~\ref{fig:muse_enviroment} does not include the core of the host galaxy, nor the southern arm. \jbu is located north of the southern distorted spiral arm of NGC~2442 and is still clearly present in NGC~2442 almost a year after maximum as seen in the white light image constructed from the datacube. The FOV ($1'\times1'$) does however include the location of SN~1999ga \citep{Pastorello2009} as well as a luminous region in the center frame. This ``\textit{Super-Bubble}'' has been noted by previous authors \citep{Pancoast_2010}, and is seen in the irregular kinematic pattern seen in the center of the FOV. Placing an age on this region is difficult, but it is likely to have formed within the last 150--250~Myr \citep{Mihos1997}. This is a spherical-looking area within the diffuse region to the south-west of the nuclear region, with a diameter of $\sim$~1.7~kpc. 

This \textit{Super-Bubble} region is in the vicinity of both \jbu and SN~1999ga \citep{Ryder2001,Pastorello2008,Pancoast_2010}. This region shows a high SFR and is bright in B-band, both signs of massive star formation. High SFR is linked with a high SN rate \citep{Botticella2012} and it is a fair assumption that the general location of this \textit{Super-Bubble} is likely to host CCSNe, as is obvious from SN~1999ga. 

The top middle panel in Fig.~\ref{fig:muse_enviroment} maps the extinction across the FOV using the Balmer Decrement \citep{Dominguez2013}. We find a value for the local extinction ($E_{B-V} < 0.45$) within 500~pc of \jbu with a similar value seen across the FOV. The top right panel in Fig.~\ref{fig:muse_enviroment} gives the velocity dispersion across the FOV. The location of \jbu lies in an area moving at $\sim-100$~\kms\ (image corrected for red-shift: z = 0.00489). The bottom left panel shows a pseudo SFR based on the extinction corrected H$\alpha$ emission \citep{Kennicutt1998}. The figure shows two bright regions of star formation, which is clear from the white light image. \jbu is situated on the outskirts of a moderate star-forming region, $\sim10^{-6}$~\msunperyr. SN~1999ga lies on the edge of the brighter star forming region. We include a metallicity map (bottom right panel) following the metallicity indicators from \citet{Dopita2016}. The full FOV yields an approximately solar environment, with the median metallicity across the field as 8.66~dex ($Z \approx 0.015$). 

\section{Bolometric evolution of \jbu}\label{sec:bolometric_evolution}

\begin{figure*}
\centering
\includegraphics[width=\textwidth]{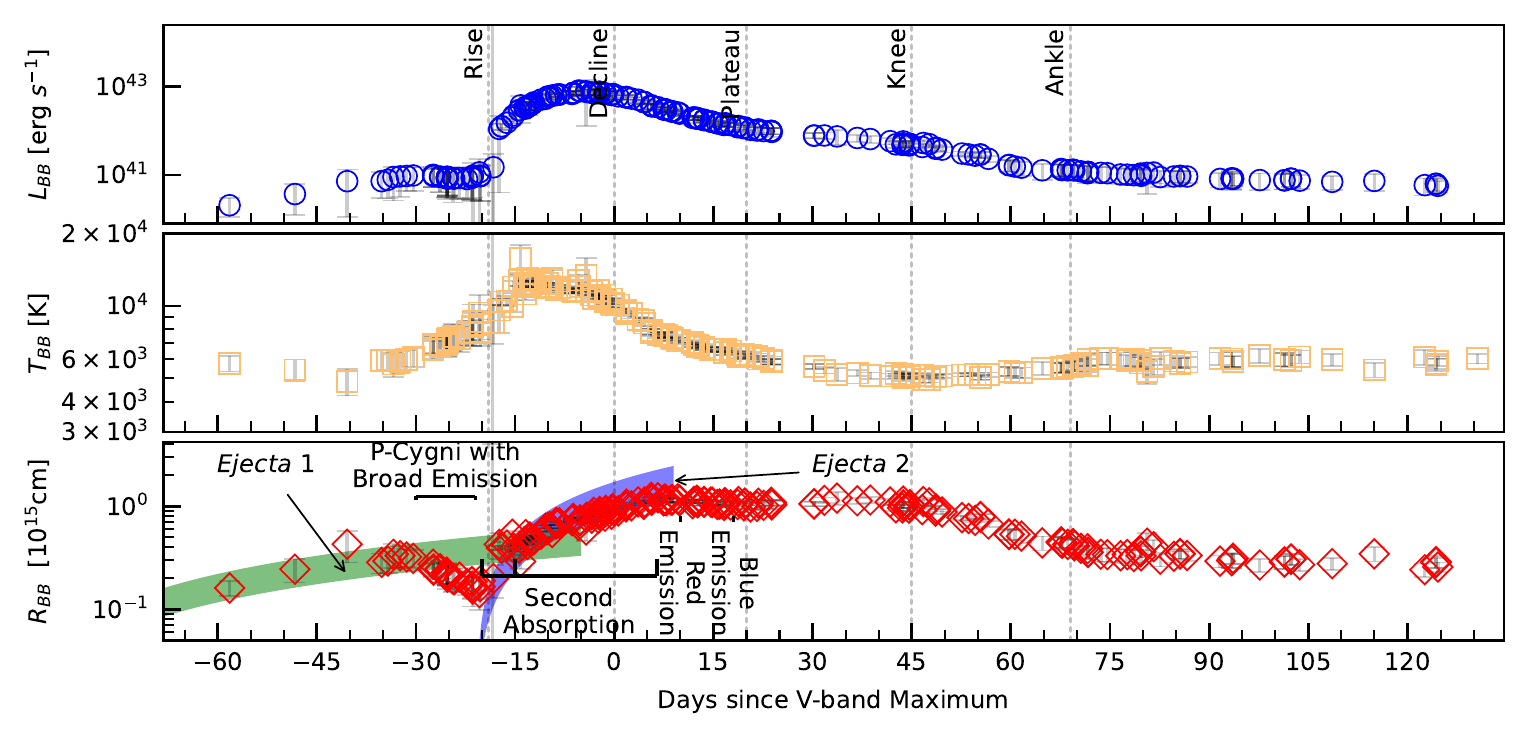}
\caption{ Blackbody luminosity (top panel), temperature (middle panel), and radius (lower panel) of \jbu calculated using \textsc{SuperBol} \citep{Nicholl_2018}. In the second and third panel, we include $T_{\rm BB}$ and $R_{\rm BB}$ fits to our optical spectra (\paperI). We include approximate epochs where specific H$\alpha$ features emerge in the $R_{\rm BB}$ panel, as discussed in the text. The green shaded region shows the linear distance travelled by the slower moving material, \textit{Ejecta 1}, causing the P Cygni absorption. The blue shaded region is the same for the faster moving material, \textit{Ejecta 2}. The lower and upper bounds for each band are bulk and max velocities respectively. }
\label{fig:bolometric_T_R}
\end{figure*}

The bolometric lightcurve for \jbu is computed using $ugiz$, $UBVR$, $JHK$, Gaia \textit{G}, $W1$ and $W2$ from WISE, as well as {\it Swift}+UVOT \textit{UVW2}, \textit{UVM2}, \textit{UVW1}, \textit{U}, \textit{B}, and \textit{V}. All calculations were carried out using \textsc{Superbol}\footnote{\url{https://github.com/mnicholl/superbol}} \citep[Version 1.7;][]{Nicholl_2018}. Effective wavelengths were taken from \citet{Fukugita1996} and zeropoint flux energies were taken from \citet{Tonry2018}, while \textsc{Superbol} was modified to also handle our WISE data. Extinction values in each filter were computed using the York Extinction Solver \citep{McCall2004}. All magnitudes were converted to $F_{\lambda}$, and interpolated where necessary to account for epochs without specific filter coverage, taking $r$-band as the reference filter. Black body fitting is performed for photometric bands that are centered on $\lambda~>~3000$~\AA\ to avoid the effects of strong line-blanketing. We also obtain a pseudo-bolometric lightcurve by directly integrating $F_{\lambda}$ using the trapezoidal rule between 0.2 and 4.5 $\upmu$m (\textit{UVW2} to \textit{W2}). We present the results of our blackbody fitting in Fig.~\ref{fig:bolometric_T_R}.

\jbu is an interacting transient showing strong emission lines. Interpreting the blackbody evolution of photometry alone may be misleading, due to the uncertainty as to whether the photometry is continuum-dominated or line-dominated. For completeness we investigate blackbody fits from our optical spectra. A black body function was fitted to the optical spectra presented in \paperI while excluding strong emission features and only fitting for $\lambda~>~3000$~\AA. We find excellent agreement with the blackbody evolution from photometric and spectroscopic data until $\sim$+125~d. After this time, our observations become strongly line-dominated and blackbody fitting becomes unreliable. 

%%%%%%%%%%%%%%%%%%%%%%%%%%%%%%
%%%%%%%%%%% RADIUS %%%%%%%%%%%
%%%%%%%%%%%%%%%%%%%%%%%%%%%%%%
\subsection{Radius and Kinematics}\label{sec:kinematices}

We show the blackbody luminosity ($L_{BB}$), radius ($R_{\rm BB}$), and temperature ($T_{\rm BB}$) fits from \textsc{Superbol} in Fig.~\ref{fig:bolometric_T_R}. The H emission for \jbu shows two distinct absorption components (see \paperI). The first component is seen in a P Cygni profile that is present up until $\sim$0.5 years after \eventB maximum. The second component is present for $\sim$1 month with respect to its first emergence, and suggests some absorbing, High Velocity (HV) material. A similar feature has been seen in other SN impostors \citep[e.g.]{Tartaglia_2016} and is common in \ip-like transients. The presence of two regions of material with different velocities lends credence to the idea that the interaction with some material during, or prior to, \eventA is the main source of energy input for \eventB \citep{Fraser13,Thone2017,Elias-Rosa2016,Benetti_2016}. 

To explore this scenario further, we assume some optically thick material causing the P Cygni absorption was ejected at $\sim$-90~d (first detection of \eventA, see \paperI), although this ejection may have occurred earlier.

Fitting a P Cygni absorption profile gives a maximum velocity of $\sim-850$~\kms, with a bulk velocity of $\sim-600$~\kms\ for the slower absorption feature seen in the Balmer lines. We refer to this material as \textit{Ejecta 1}. The higher velocity absorption (which we refer to as \textit{Ejecta 2}) has a maximum velocity from the blue edge of the line of $\sim-10000$~\kms\ with the bulk of the material at $\sim-4500$~\kms. Using these velocities, we attempt to constrain ejection/collision times.

The ejection epoch for the material causing the second high velocity absorption component is open to debate. There is no evidence for this additional absorption in optical spectra at $-24$~d and it is only seen on $-15$~d. Under the presumption that we do not see this shell of material (i.e \textit{Ejecta 2}) until it interacts with the pre-existing material or until it is no longer occulted by an existing photosphere, we find that a shell moving at $\sim$4500~\kms\ for $\sim$3~days can reach the distance of $R_{\rm BB}\sim$\SI{0.1e15}{cm}. We include the distance travelled by \textit{Ejecta 2} in Fig.~\ref{fig:bolometric_T_R} as a blue band. We can constrain the ejection date of this HV material to $\sim$21 days before maximum light with the collision date (when \textit{Ejecta 2} catches up to \textit{Ejecta 1}) at $\sim$19 days before maximum light.

We draw attention to the blackbody evolution over the period $-19$~d to $-13$~d. During this timeframe we see an inflection between the decline of \eventA and the rise of \eventB. Although we have low-cadence coverage during \eventA, the distance travelled by \textit{Ejecta 1} follows $R_{\rm BB}$ quite well during \eventA. $R_{\rm BB}$ then contracts slightly beginning around $-30$~d to a minimum at $-19$~d. At $\sim-19$~d $R_{\rm BB}$ increases at a velocity similar to the velocity profile of \textit{Ejecta 2}. This implies that the blackbody radius now follows this material, which is likely \textit{Ejecta 2} with additional material swept-up from \textit{Ejecta 1} and some CSM material. While this is undoubtedly a simplified picture, it appear that \jbu is potentially consistent with two successive eruptions (either non-terminal or a CCSN) where the collision of ejecta powers the luminosity of \eventB.

We initially find $T_{\rm BB}$ at $\sim5700$~K, which is roughly constant up until $\sim-30$~d. $T_{\rm BB}$ evolves exponentially from 6000~K at $-30$~d to 12000~K at $-12$~d. After \eventB maximum (marked as \textit{Decline} in Fig \ref{fig:bolometric_T_R}), $T_{\rm BB}$ cools to $\sim5100$~K at the \textit{Knee} epoch and slightly increase to $\sim6000$~K at the beginning of the \textit{Ankle} epoch.

It is important to note that we see both components in spectra during the first month of \eventB. Additionally, the FWHM and velocity offset does not significantly evolve during the first few months (see \paperI). 

This is likely due to \textit{Ejecta 1} or the CSM or both being highly asymmetric. We are motivated by the spectral evolution of the H$\alpha$ profile, the evolution of $R_{\rm BB}$ and the degenerate appearance of the H$\alpha$ emission lines in \ip-like objects, see \paperI for further discussion. If \textit{Ejecta 2} is spherically symmetric (e.g. possibly a CCSNe), some material of \textit{Ejecta 2} would not interact with \textit{Ejecta 1} and expand freely along the lower density regions. 

We include labels indicating when certain spectral components appear in the H$\alpha$ in Fig.~\ref{fig:bolometric_T_R}, bottom panel. We see that the HV blue absorption feature coincides with the evolution of \textit{Ejecta 2}; this absorption is clearly seen at $-18$~d and is detected until $+5$~d with fitting-model dependent tentative detections up to $+10$~d. This second absorption component appears during the rise in $R_{\rm BB}$ during \eventB and vanishes when $R_{\rm BB}$ reaches its maximum at $\sim$+7~d. At $\sim$+9~d, $R_{\rm BB}$ remains at a constant value and we see the emergence of a broad, red shoulder emission in H$\alpha$ at $\sim~1400$~\kms, FWHM~$\sim~4000$~\kms. This may follow material expanding at $\sim1400$~\kms, a receding photosphere or both. Several days later the blue emission feature appears in H$\alpha$ and remains until late times. At $+18$d this blue emission is centered at $\sim~-2400$~\kms\ with FWHM~$\sim$3800~\kms. 

Photons from the interaction site between \textit{Ejecta 2} and \textit{Ejecta 1}/CSM may be diffusing outwards at this epoch. We see that the red shoulder emission only appears after $R_{\rm BB}$ reaches its maximum values, shortly followed by the blue shoulder emission a week later. This leads to our conclusion that \textit{Ejecta 1} is partially asymmetric and when \textit{Ejecta 2} collides with it, \textit{Ejecta 1} is partially engulfed. The interaction between these two shells then becomes apparent at $\sim$+7~d when the asymmetric emission features are clearly seen in H$\alpha$.

The $R_{\rm BB}$ peaks at $1.2\times10^{15}$~cm, at $\sim1$~week after \eventB maximum and remains roughly constant until the \textit{Knee} phase. Thereafter, there is a drop of $\sim-5\times10^{13}$~cm/day until the beginning of the \textit{Ankle} phase. $R_{\rm BB}$ remains roughly constant at $\sim0.3\times10^{15}$~cm up until the seasonal gap begins at $+140$~d. This epoch coincides with a narrowing of both red and blue emission features and an increase in Equivalent Width (EW) of both components. This may represent a time when opacities drop significantly and there is less photon scattering. Using this collision scenario as the dominant energy input for this transient, we will explore the necessary energy budget in Sect.~\ref{sec:power}. Using the evolution of $R_{\rm BB}$ we can better understand the nature of the explosion of \jbu, and we will further discuss this in Sect.~\ref{sec:discuss_geometry}. 

%%%%%%%%%%%%%%%%%%%%%%%%%%%%%%%%%%%%%%
%%%%%%%%%%%% POWERING JBU %%%%%%%%%%%%
%%%%%%%%%%%%%%%%%%%%%%%%%%%%%%%%%%%%%%

\section{Powering \jbu}\label{sec:power}

The nature of the energy input of \jbu and \ip-like transients is debated. If \jbu is indeed a CCSN then this energy comes from an imploding iron core and the early lightcurve is powered by the fast moving SN ejecta material. Ejecta interacting with a dense CSM can power the lightcurve for many years \citep[see][and references therein]{Fraser2020}. If the transient is a CCSN, after the ejecta expands and cools, the late time lightcurve is powered by the radioactive decay of $^{56}$Ni. We discuss the possible presence of $^{56}$Ni in Sect.~\ref{sec:ni}.

If \jbu is a CCSN, then it is spectroscopically classed as Type IIn, meaning we see strong signs of interaction with a dense, slow-moving CSM. We discuss the energy input from ejecta/CSM interaction in Sect.~\ref{sec:CSM_eject_interaction}.

\subsection{\texorpdfstring{\textsuperscript{56}}\ Ni mass}\label{sec:ni}

A product of CCSNe is explosively synthesised radioactive $^{56}$Ni, whose decay can power the late time lightcurve of H-rich supernovae, after the hydrogen ejecta have fully recombined and any additional interaction has stopped. \cite{Anderson2019} find that for H-rich, Type II SNe, the median value for the amount of $^{56}$Ni synthesised is 0.032~\msun. We show our attempt to fit for a nickel decay tail in Fig.~\ref{fig:bolometric_L} (green dashed line). We find that the pseudo-bolometric lightcurve shows a decay that is consistent with that of radioactive nickel decay during the \textit{Ankle} stage. 

Determining an explosion epoch for \jbu is contentious. The transient is clearly detected at -90~d in VLT+FORS2 imaging. We determine in Sect.~\ref{sec:kinematices} that a second eruption (that may represent a genuine CCSN) occurred at $\sim$-21~d. Using eq. 6 from \citet{Nadyozhin_2003}, and taking the explosion epoch as $\sim$-90d, we find a value of $M_{Ni}\leq 0.033$~\msun\ and taking $\sim-21$~d we find a value of $M_{Ni}\leq0.016$~\msun. Following the arguments made in Sect.~\ref{sec:kinematices} we will take the latter explosion date as the more plausible motivated by the apparent second eruption at $-21$~d ( this is the explosion epoch typically assumed in the literature ), indicating a potential CCSN.

This limit on $^{56}$Ni is consistent with other \ip-like transients. However, it is clear that during this time there is still on-going CSM-interaction, as demonstrated from the multi-component H$\alpha$ profile in \paperI, and as such, this value should be considered a conservative upper limit, assuming any $^{56}$Ni is produced at all.

\subsection{CSM-ejecta interaction}\label{sec:CSM_eject_interaction}

\begin{figure*}
\centering
\includegraphics[width=\textwidth]{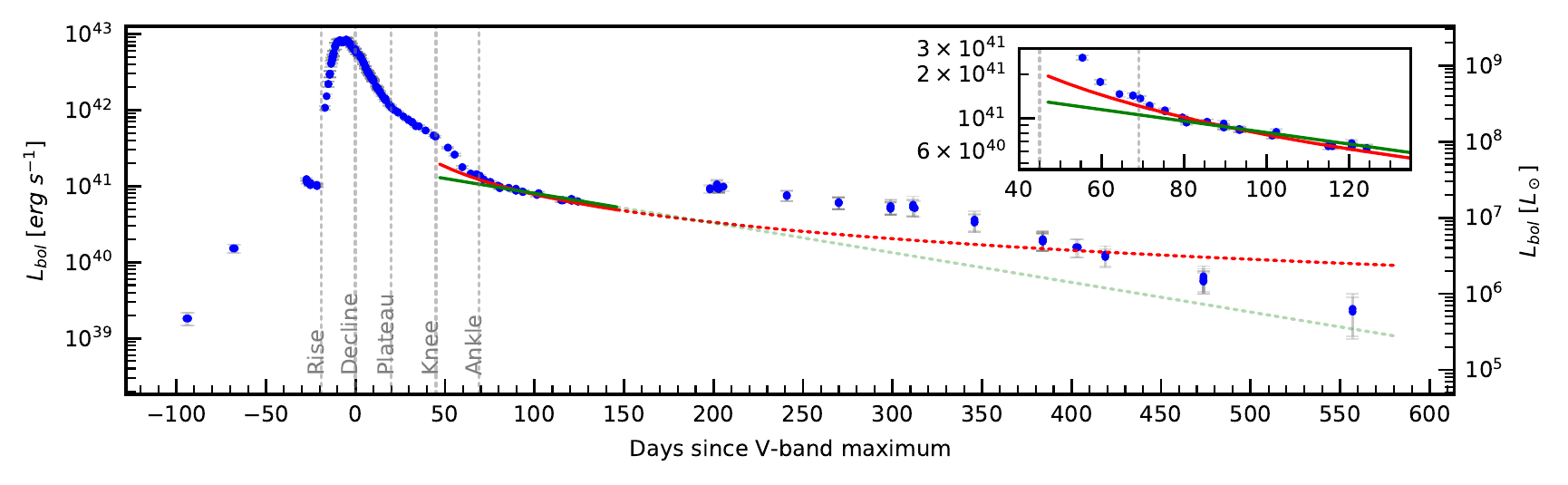}
\caption{Pseudo-bolometric luminosity of \jbu using \textsc{SuperBol} \protect\citep{Nicholl_2018}. We include the luminosity shock function (Eq. \ref{eq:luminosity_shock_eq}; solid red line) and a radioactive decay tail fit (green solid line). Both functions are extrapolated until the end of observations (dashed lines). Both function are fitted to the post-ankle stage and we include a zoom-in of this area in the top right. We find a $^{56}$Ni mass of 0.016~\msun\ (assuming the SN explosion date as $-21$~d) and $\dot{M}$ is 0.05~\msunperyr for Eq. \ref{eq:luminosity_shock_eq}.}
\label{fig:bolometric_L}
\end{figure*}

A previously explored scenario for the double-peaked lightcurve of \ip-like objects is \eventA represents a low energy eruption from the progenitor star and \eventB is powered by the interaction between the ejecta from this eruption, and some pre-existing CSM that was ejected in the preceding years \citep{Mauerhan13,Fraser13,Thone2017}.

We measure the radiated energy released from \eventA ($-90$~d to $-21$~d) as \SI{3.15e+47}{erg} and the energy from \eventB ($-21$~d to $+450$~d) as $\sim$~\SI{1.81e+49}{erg}. \citet{Fraser13} find a similar value for \ip at $\sim$~\SI{1.8e49}{erg}.

If we assume that \eventA is a symmetric explosion \citep[similar to that proposed in ][]{Mauerhan_2014} we can approximate it using an \textit{Arnett Model} \citep{arnett1996}. Taking the diffusion timescale for a photon to be $t_d \approx \frac{L^2}{D}$, where D is a diffusion coefficient with $ D \approx \lambda c~=~\frac{c}{\rho~\kappa}$. Assuming that \eventA corresponds to the adiabatic expansion of a photosphere, and assuming $L \approx R$, we can describe the diffusion timescale as:

\begin{equation}\label{eq:mass_ej}
 \tau_d = \left(\frac{3~\kappa~M_{ej}}{4~\pi~c~v_{ej}}\right)^{1/2},
\end{equation}

\noindent by substituting in $R \approx \tau_{d} \times v_{sh}$ and $ \rho \approx \frac{3~M_{ej}}{4~\pi~R^3}$, where $M_{ej}$ and $v_{ej}$ are the ejecta mass and velocity respectively, $\kappa$ is the opacity of the ejecta, and $c$ is the speed of light. We take the rise time in \textit{r}-band of \eventA to be similar to the diffusion time, and we get a value of $\sim$60~days. We assume the P Cygni minima follows this dense material ejected prior to, or during, the beginning of \eventA, as suggested by \citet{Thone2017} for \bh. Using Eq. \ref{eq:mass_ej} and taking $v_{ej}\approx700$~\kms\ and assuming a mean opacity of $\kappa$ = 0.34 $cm^2 g^{-1}$ (assuming $e^{-}$ scattering dominates in the H-rich ejecta) we find $M_{ej}$ for the \eventA is $\sim$0.35~\msun\ giving a kinetic energy of $\sim$~\SI{1.7e48}{erg}. 

This value is a factor 10 less than is required to power \eventB. This is a crude approximation as we invoke spherical symmetry. To fully investigate the mass of \textit{Ejecta 1}, detailed hydrodynamic simulations are needed \citep[e.g.][]{Vlasis2016,Suzuki2019}, which are beyond the work presented in this paper. 

Assuming free expansion, the constrained ejection times, and velocities for our multiple shell models given in Sect.~\ref{sec:kinematices}, the beginning of \eventB coincides with material from both shells being at the same location $R_{\rm BB}\approx$ \SI{0.7e15}{cm} (Fig.~\ref{fig:bolometric_T_R}). This suggests that \eventB is powered from the collision at $\sim-19$~d of \textit{Ejecta 2} which interacts with the slower moving material ejected at the beginning of \eventA, (\textit{Ejecta 1}).

It is difficult to measure the mass of \textit{Ejecta 2}. If we assume that \eventB is powered solely by CSM interactions, we calculate that $M_{ej}~\sim~0.37$~\msun\ travelling at $\sim$~5000~\kms\ can account for the energy seen, while allowing for an extremely low porosity (or overlapping surface area between the colliding material) of $10\%$. This value will change depending on the opening angle of the interaction site, as explored in disk interaction models \citep{Vlasis2016,Suzuki2019,Kurfurst2020}. 

Even with this conservative estimate, our values of $M_{ej}$ are much lower than those seen in CCSNe or \etacar. However, extremely low porosity (e.g. 1$\%$) would allow for a few~\msun\ of ejected material if we assume no input to the lightcurve from radioactive decay.

Although observed after peak luminosity, both SN~2013L and SN~2010jl showed a plateau phase after maximum light \citep{Ofek2014,Taddia2020}. This trend is discussed by \citet{Chevalier2011}; SN ejecta interacting with a dense mass loss region can form a plateau in luminosity lasting the duration of the shock interaction, and ending when the entire interaction material is shocked. As the photon mean free path increases with the geometric expansion of the CSM, the innermost regions of the interaction are revealed. This was suggested to explain the double-peaked spectral profiles of SN~2010ij \citep{Ofek2014}, SN~2013L \citep{Taddia2020}, and iPTF14hls \citep{Andrews2018,Sollerman2019,Moriya2020} at late times. We use the emergence of the blue emission feature and the decrease of the peak velocity offset as a proxy for the shock front. We discuss the evolution of this feature in \paperI. We fit a declining power law function to the peak velocity of the blue emission from +20 to +120~days which is well fit by:

\begin{equation}\label{eq:power_law}
 v_{blue}(t) \approx (1375 \pm 25) \times \left(\frac{t}{100d}\right)^{-0.40 \pm 0.03} \mathrm{km~s^{-1}},
\end{equation}

\noindent Both red and blue emission components follow Eq. \ref{eq:power_law} well (the red component has a different normalisation constant) up until the seasonal gap ($+140$~days). After that both components maintain at a higher velocity and coast at $\sim\pm$1300~\kms\ up until the end of our spectroscopic observations ($+575$~days), see \paperI. Under the assumption of steady-state mass loss, the luminosity from CSM-shock interaction can be described by:

\begin{equation}\label{eq:luminosity_shock_eq}
 L_{sh} = \epsilon~\frac{1}{2}~\frac{\dot{M}}{v_{wind}}~v_{ej}^3,
\end{equation}

\noindent where $L_{sh}$ is the luminosity from CSM-ejecta interaction, $\epsilon$ is the conversion efficiency from kinetic to thermal energy (taken to be 50\%, typical of Type IIn SNe \citep{Smith_2017}), $v_{ej}$ is the ejecta velocity, which is set to Eq. \ref{eq:power_law}, and $v_{wind}$ is the wind velocity. We fit Eq. \ref{eq:luminosity_shock_eq} to our bolometric lightcurve during the period from the \textit{Knee} stage up until the beginning of the seasonal gap. Fitting to this time-frame gives an upper limit for $\dot{M}\approx0.05$~\msunperyr, if we assume an LBV wind with $v_{wind}\approx250$~\kms\ (we find a similar value for $v_{wind}$ from our earliest H$\alpha$ profile). Setting $v_{wind}\approx700$~\kms, the value of the P Cygni minima, we obtain $\dot{M}\approx0.14~$~\msunperyr.

We base the above calculations on the assumption that the luminosity between +70~d to +140~d is shock-CSM interaction dominated, with no other major contributing energy source i.e. no major contribution from radioactive decay. If \jbu is surrounded by a dense, disk-like CSM, the assumption that this phase is interaction dominated is motivated by models \citep[e.g. Fig. 11 from][]{Vlasis2016}. These models show a similar lightcurve shape to \jbu, including a tail resembling radioactive $^{56}$Ni decay at $\sim$+80~days past maximum brightness (these models assume no $^{56}$Ni). Symmetric ejecta and disk interaction models show that the energy input at the \textit{Knee} stage is dominated by this ejecta-disk interaction. We will return to the possibility of disk-like CSM in Sect.\ref{sec:discuss_geometry}.

After the seasonal gap ($+140$~days), the velocity of the red/blue emission does not follow Eq. \ref{eq:power_law} and the bolometric luminosity does not follow Eq. \ref{eq:luminosity_shock_eq}. At this point the lightcurve has increased in brightness, which is clearly seen in Fig.~\ref{fig:bolometric_L}. However by $\sim400$~d, $L_{bol}$ fades below the extrapolated value from Eq. \ref{eq:luminosity_shock_eq}.

After the seasonal gap both red and blue emission lines have similar FWHM, $\sim$1500~\kms, with the red emission having a slightly larger width, but converging to the FWHM of the blue component at $\sim$400~d. If the red/blue emission follows the shock interaction, this suggest an increased velocity of the shock front. Conserving mass flux in the shock we have $\rho_1~v_1 = \rho_2~v_2$ where subscript 1,2 represent the post- and pre- shock regions respectively. If the shock transverses to a lower density CSM environment this can account for the increased velocity seen. This might indicate that the shock has now reached a lower density environment, perhaps created by the series of outbursts in the years prior. However, it is not obvious how interaction with a less dense region of CSM would account for the increased luminosity as well as the increased strength of \ion{He}{I} emission lines (also seen in SN~1996al; \citealp{Benetti_2016}) at this time (\paperI). 

%%%%%%%%%%%%%%%%%%%%%%%%%%%%%%
%%%%%%%% DISCUSSION %%%%%%%%%%
%%%%%%%%%%%%%%%%%%%%%%%%%%%%%%
\section{Discussion}\label{sec:discuss}

In the following section we will discuss the nature of \jbu. There is much debate as to the nature \ip-like objects \citep{Pastorello2008,atel4412,Fraser13,Smith2014,Margutti2014,Graham2014,Pastorello2019}. Any scenario for \jbu or \ip-like transients needs to account for all of the following points:

\begin{enumerate}[label=\bfseries \arabic*:,leftmargin=*,labelindent=1em]

    \item Outbursts reaching an absolute magnitude of $M_r\sim-11\pm2$~mag seen in the historic lightcurve of the transient. 
    
    \item A faint event, reaching an absolute magnitude of $M_r\sim-13\pm2$~mag.
    
    \item An second event a few weeks later, reaching an absolute magnitude of $M_r\sim-18.5\pm0.5$~mag and ejecting material with velocities up to $\sim~10,000$~\kms.
    
    \item Ejected $^{56}$Ni mass of $\lesssim$~0.02~\msun.
    
    \item No directly observed synthesized material, either from explosive nucleosynthesis or late-stage stellar evolution.
    
\end{enumerate}

A possible addition to this list is double-peaked emission lines. This is seen in the majority of \ip-like transients although, ironically, not \ip itself.

We address the probable progenitor in Sect.~\ref{sec:discuss_progenitor}. Using our high cadence multi-chromatic photometry presented in \paperI, and the bolometric evolution from Sect.~\ref{sec:bolometric_evolution}, we will present a likely explosion model and circumstellar (CS) geometry for \jbu, that can be extrapolated to other \ip-like transients in Sect.~\ref{sec:discuss_geometry}. We will discuss the validity of a CCSN scenario in Sect.~\ref{sec:discuss_snec} and Sect.~\ref{sec:discuss_CCSN}, and the possibility of the progenitor being in an interacting binary system in Sect.~\ref{sec:discuss_binary}.

\subsection{The Progenitor of \jbu and \ip-like transients }\label{sec:discuss_progenitor}

The events of \ip-like transients may represent a critical step in the late time evolution of massive stars. A dramatic increase in luminosity allows for supper-Eddington winds and high mass-loss rates, however the mechanism resulting in these outburst is unknown. Observations of shock features in the Homunculus Nebula around \etacar may even point to explosive mass loss. Furthermore, in the classical picture, LBVs should not be SN progenitors as they have just transitioned to the He-core burning stage in their core 

It is generally thought that \ip-like transients arise from very massive stars \citep{Foley2011,Pastorello2013,Fraser13,smith13,Fraser2015,Smith2016,Elias-Rosa2016,Pastorello2019}. The progenitor of \ip is thought to be a 60--80~\msun\ LBV from pre-explosion images \citep{smith10,Foley2011}. However, this was measured in a single band only, which may be strongly affected by H$\alpha$ emission. As shown in Fig.~\ref{fig:hst_SED}, the bright contribution of H$\alpha$ in \textit{F350LP} will provide misleading SED fitting results. While LBVs experience erratic mass loss as they undergo a short transition from O-Type to the WR stars, \jbu appears to be too low mass ($\sim$22~\msun) to be consistent with the \ip progenitor. We note that this relatively low mass was found while taking into account the effect of H$\alpha$ emission on the SED. 

Our analysis on the progenitor mass for \jbu is the most secure for any \ip-like transient in literature, as it is based on a broad optical to NIR SED, as well as on the local neighbourhood. From our SED fitting to the early 2016 {\it HST} data, we find the color of the progenitor is consistent with a yellow hyper-giant. Using {\sc dusty} modelling and matching the output spectra to these colour values we find values for L and T which are consistent with a mass of single star of 22--25 \msun, consistent with the results from \citetalias{Kilpatrick2018}. Moreover, the local environment, which can be be assumed to be composed of a similar stellar population, demonstrates that we can effectively rule out a very young population (expected for a 60--80~\msun\ star). 

In order to explore the progenitor further, we turn to a grid of stellar models created with the {\sc BPASS} code. The {\sc BPASS} stellar model library contains the time varying properties of over 250,000 star systems for a grid of initial parameters and a population containing a realistic fraction of binary and single star systems \citep{Eldridge2017, Stanway2018}. Using {\tt hoki}\footnote{\url{https://github.com/HeloiseS/hoki}} \citep{stevance2020}, we searched for models matching the observed temperature and luminosity of the progenitor of \jbu, considering both the possibility of a terminal core-collapse supernova and of a non-terminal event. 

For the CCSN (non-terminal explosion) scenario we find 12 (1668) matching stellar models, and 0 (3) of these models correspond to single star systems.

\begin{figure}
\includegraphics[width=\columnwidth]{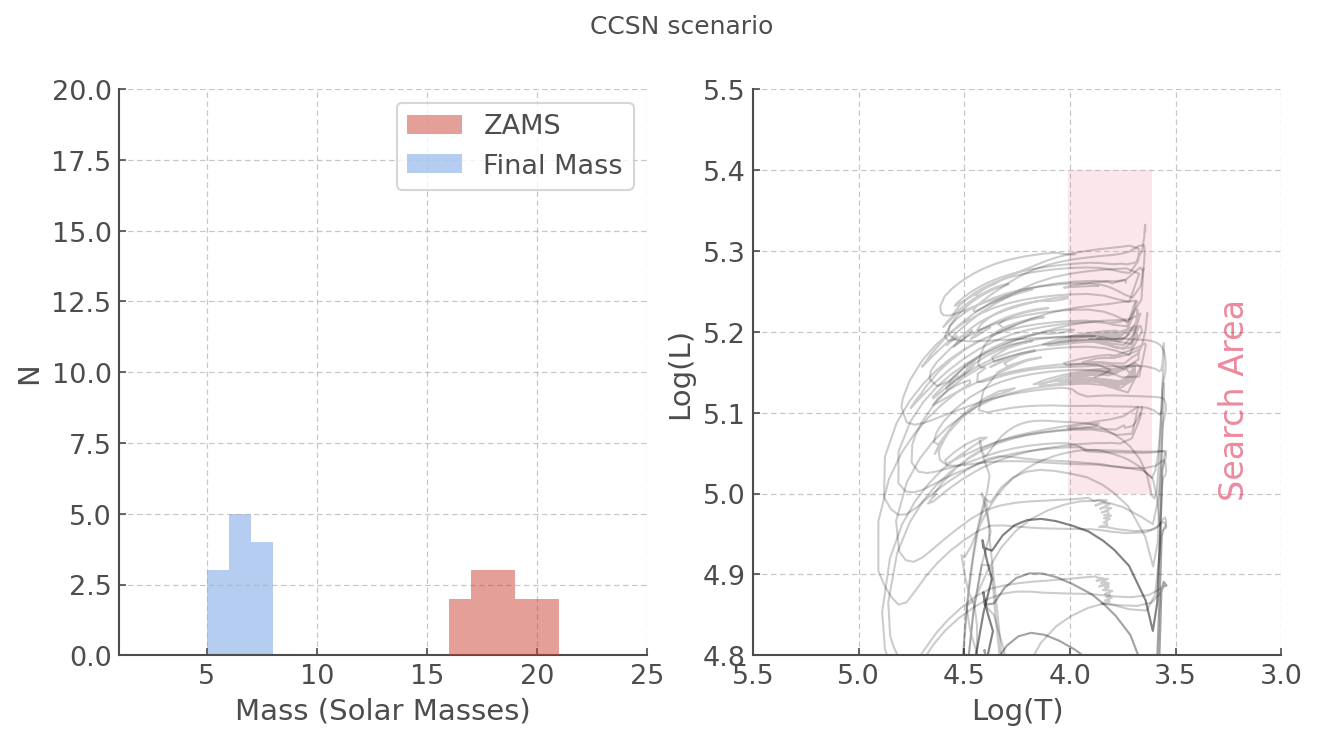}
\includegraphics[width=\columnwidth]{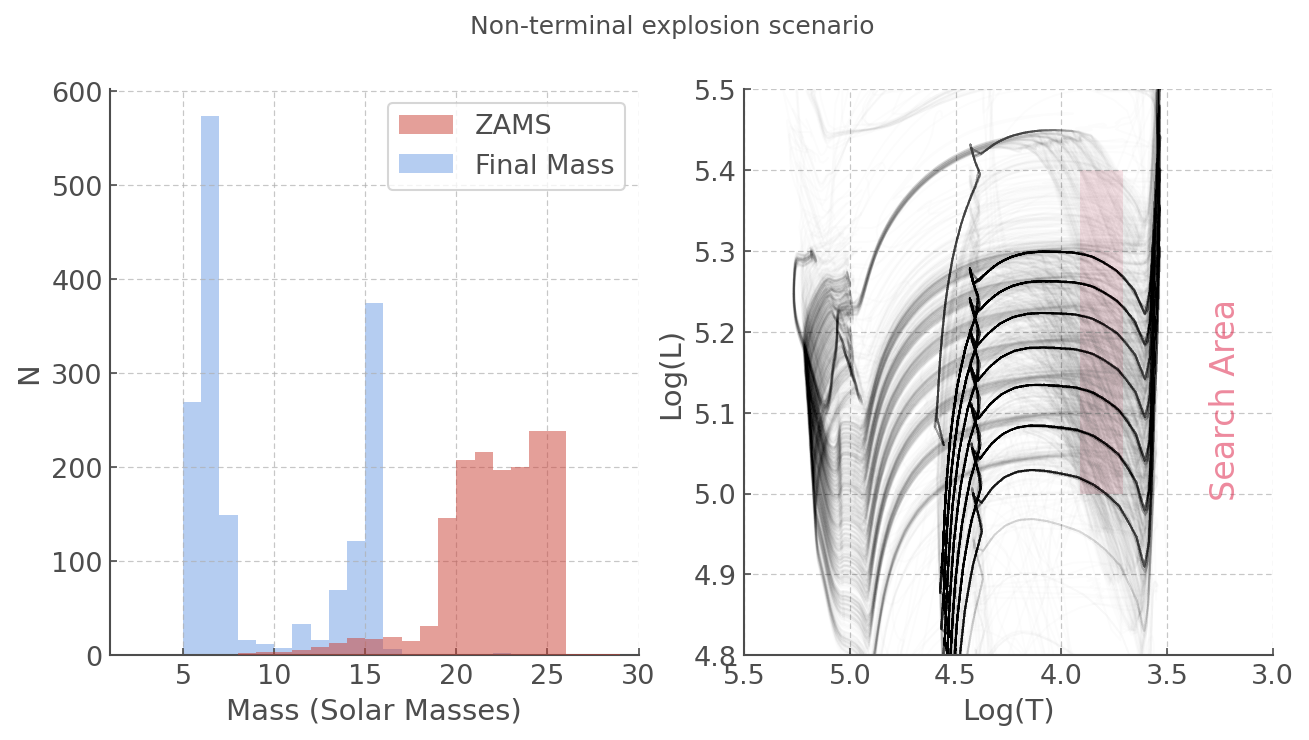}
\caption{Mass distributions (ZAMS and final) and evolutionary tracks of {\sc BPASS} models matching the L and T derived from {\sc dusty} modelling. Upper (lower) panel shows the CCSN (Non-terminal) scenario. Each evolutionary track is plotted at a low transparency and therefore the lighter the tracks, the rarer they are in our matches. We mark the search area in T and L from Sect.~\ref{sec:progenitor} in each HR diagram.
}
\label{fig:HRD_bpass}
\end{figure}

The ZAMS and final mass distributions, as well as the evolutionary tracks for both interpretations, are presented in Fig. \ref{fig:HRD_bpass}. We can evaluate the mean and standard deviation for the two scenarios: $\overline{{\rm M}}=12.3~$\msun, $\sigma = 1.9$~\msun\ and $\overline{{\rm M}}=22$~\msun, $\sigma= 3.4$~\msun\ for CCSN and non-terminal explosion cases, respectively.

% $\log(\tau_{\rm life})={7.31^{+0.10}_{-0.12}}$ yrs and $\log(\tau_{\rm life})={7.13^{+0.14}_{-0.22}}$ yrs
% 11.106 +/ 0.471 -/ 1.103

We find a mean lifetime of $7.3^{+0.1}_{-.1}$~Myr and $7.0^{+0.1}_{-0.1}$~Myr for the CCSN and non-terminal explosion scenarios, respectively. Very massive stellar progenitors (e.g. classical LBVs with $>50$~\msun) are confidently excluded for \jbu. 

There are numerous suggestions in the literature that LBVs can be the direct progenitors of CCSNe \citep[e.g.][]{Trundle2008,Dwarkadas_2011,Smith_2014,Humphreys_2016,Ustamujic2021}. It has been suggested that the LBV phenomenon may occur in stars with initial masses as low as $20 - 25$~\msun, particularly when rotation is included in models \citep{Groh2013}. Such LBVs may appear similar to F-type yellow super giants during their eruptive stage \citep{Kilpatrick2018,Humphreys_2016}. So, it is possible that the progenitor of \jbu is a low-mass LBV. However, we still require a high mass-loss of $\sim$0.05 \msunperyr\ to explain the lightcurve of \jbu. This is not dissimilar to the mass loss rate of \etacar during its Great Eruption ($\sim0.1$ \msunperyr; \citealp{Davidson97}), but it remains unknown whether {\it lower} mass LBVs can sustain such a high mass loss rate.

\subsection{Geometry of \jbu and \ip-like transients}\label{sec:discuss_geometry}

An interesting problem to solve with the CCSN scenario is that of the presence and geometry of the CSM, as discussed in Sect.~\ref{sec:bolometric_evolution}. The LBV-type winds invoked in Sect.~\ref{sec:CSM_eject_interaction} do not apply to lower mass progenitors; indeed we find an average mass-loss rate over the last 1~Myr of $\log(\dot{\rm M})= -5.4^{+0.2}_{-0.8}$ \msunperyr\ and $\log(\dot{\rm M})= -4.9^{+0.2}_{-0.3}$ \msunperyr\ for the CCSN and non-terminal scenario, respectively.

One can sustain a dense CSM even with a low mass loss rate provided the wind velocity is sufficiently small. Using $\log(\dot{\rm M})$/$v_{wind}$ as a proxy for wind density, we compare the average ratio found in our models to that assumed in Sect~\ref{sec:CSM_eject_interaction}. We find that for both sets of  progenitor models $\dot{\rm M}$/$v_{wind}\approx10^{-6}$, compared to a value of $\approx10^{-4}$ found for \jbu. Thus, we can confidently assert that steady winds are not able to create the CSM observed in \jbu.

The alternative is episodic mass loss resulting from Roche lobe overflow (RLOF) or common envelope evolution (CEE). We examined the CCSN progenitor models found in {\sc BPASS} and find that 3 models are in a CEE phase at the time of CCSN explosion; furthermore we find another 2 undergoing mass transfer. Similarly, for the non-terminal models we find that 937 models are in the CEE phase and 501 are undergoing stable mass transfer at the point where they match the observed L and T of the \jbu progenitor. Consequently, the {\sc BPASS} models reveal that the peculiar combination of properties and environment of \jbu can be explained by binary interactions. 

A radially confined, dense, disk-like CS environment has been suggested for \ip-like transients \citep{Smith2014,Margutti2014,Levesque2014,Margutti2014,Fraser2015,Benetti_2016,Tartaglia2016,Pastorello_2017,Andrews2018} as well as other Type IIn SNe \citep{vanDyk1993,Benetti_2000,Stritzinger_2012,Benetti_2016,Andrews2017,Nyholm2017} and super-luminous supernovae (SLSNe) \citep{Metzger_2010,Vlasis2016}.

Double-peaked line profiles are signs of an asymmetric environments such as a disk, rings, or bipolar outflows cause by an asymmetric explosion. This is similar to the presence of double-peaked H$\alpha$ (and other emission lines) originating from an accretion disks in active galactic nuclei (e.g. \citealp{Shapovalova2004}) as well as double-peaked emission from Be/shell stars (e.g. \citealp{Silaj2014}), although their formation and powering mechanism are extremely different. We show in \paperI that \jbu and other \ip-like objects show a degree of degeneracy in the appearance of their H$\alpha$ profiles, which may be explained with a simple viewing angle effect.

We suggest that \jbu has underwent a series of eruptions, such as has been suggested for \etacar \citep[see review by][]{smith2009} and \ip \citep{Mauerhan_2014,Levesque2014,Margutti2014,Reilly2017}, and a significant portion, if not all, of the explosion energy is a result of a ejecta-ejecta or ejecta-CSM interaction, which dominates around a month after maximum light. It is uncertain whether any of these eruptions emanate from core-collapse.

Recently, several groups have modelled the interaction of ejecta with aspherical CSM \citep{Vlasis2016,McDowell2018,Suzuki2019,Kurfurst2020,Nagao2020}. \citet{Vlasis2016} has modeled the lightcurve evolution of a spherically symmetric eject colliding with a disk-like CSM. We find that similarities between these models and \jbu. One important feature is after $\sim$~+80days these models seem to follow a decay similar to that expected from $^{56}$Ni. The energy source at this time is solely powered from CSM interaction and not from radioactive decay. However, these models cannot explain the increased brightness in \jbu after the seasonal gap, although this likely reflects a clumpy CSM and would require fine-tuning of the CSM density profile.

Models by \citet{Kurfurst2020} have modeled ejecta interaction with aspherical CSM for a range of viewing angles \citep[Model A and Fig.~12 in][]{Kurfurst2020} demonstrating a clear viewing angle degeneracy, with looking down through the plane of the CSM showing the greatest ``\textit{double-peaked}"-ness and looking through the material showing the least (i.e singularly peaked emission lines). This can naturally explain the variations in H$\alpha$ appearance found amongst \ip-like transients (see \paperI).

% For contrast, SN 1987A shows an expanding ring system with a velocity of only $\sim10$~\kms \citep{Crotts1991}. \etacar may represent a more useful guide \citep[see][and references therein]{smith2013}. More specifically, analysis by \citet{smith2006} suggest that the bipolar lobes and equatorial skirt seen in {\it HST} images must have been ejected around the same time, effectively ruling out a pre-existing asymmetric CSM constrained to the equator. Another, albeit highly speculative possibility is that a binary merger immediately prior to (or triggering the events seen in \jbu) could lead to the necessary disk-like ejecta \citep{Soker_2013,Pastorello_2019}.

For \ip-like transients, there is some discrepancy as to the eruption epoch of this asymmetric structure, with some authors suggesting this material was ejected close to/during  \eventA \citep[e.g.][]{Margutti2014,Tartaglia_2016,Thone2017} whereas some authors speculative the disk has been ejected much longer before \citep[e.g.][]{Mauerhan_2013,Mauerhan_2014}. This is difficult to understand without specific stellar evolutionary models. 

\begin{figure*}
\centering
\includegraphics[width=\textwidth]{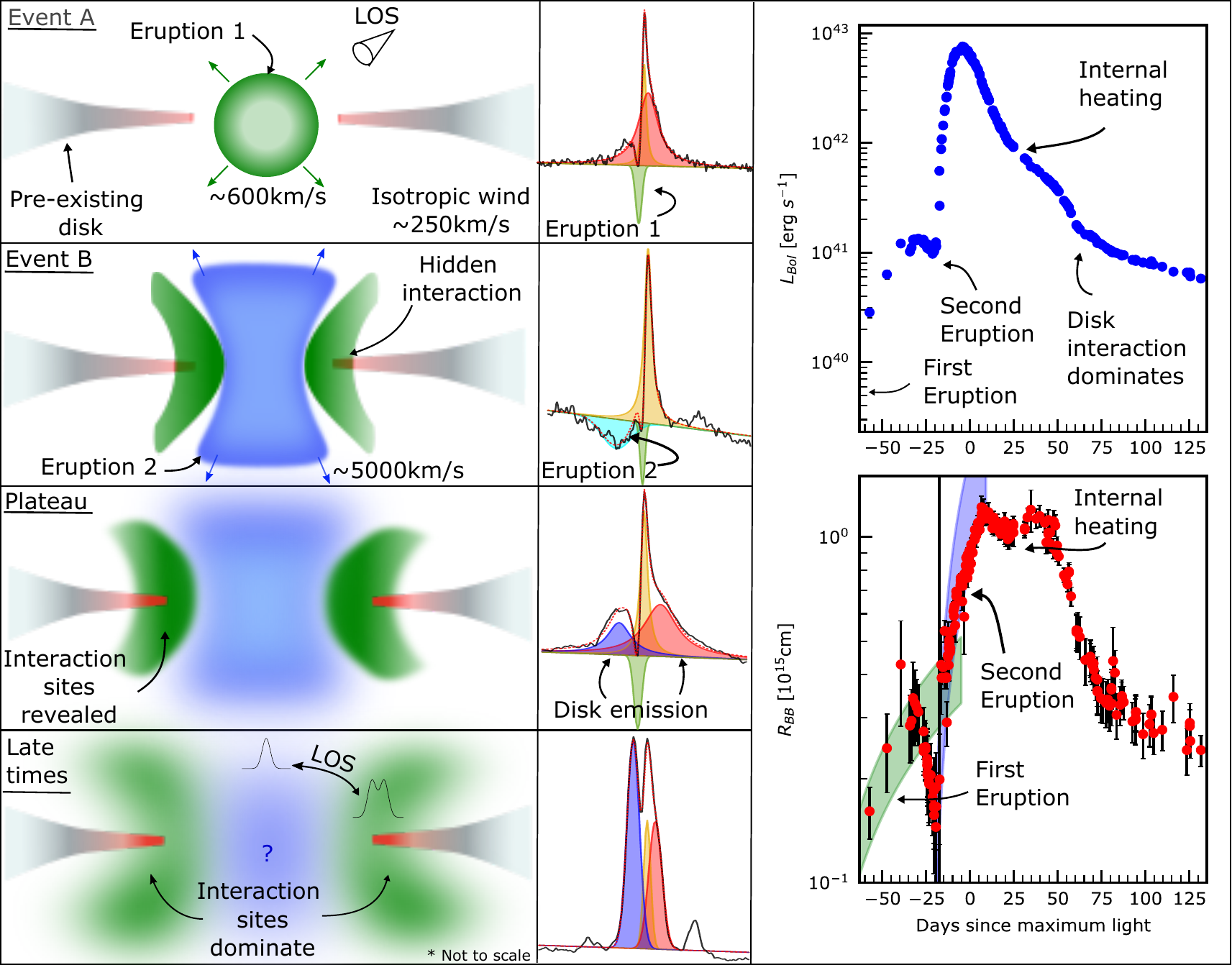}
\caption{Toy model depicting possible geometry and explosion scenario for \jbu. This diagram illustrates the discussion in Sect.~\ref{sec:kinematices} and Sect.~\ref{sec:CSM_eject_interaction}. Left panels show a simplified illustration of the CS environment around the progenitor at specific epochs. Middle column shows H$\alpha$ profile at corresponding epochs. The upper right panel shows the bolometric luminosity and the lower right shows the blackbody radius. We include the distance travelled by \textit{Ejecta 1} (green shaded region) and \textit{Ejecta 2} (blue shaded region). \eventA begins with an eruption {of ejecta 1} which originates from the progenitor system. The eruption and expansion of \textit{Ejecta 1} is seen in $L_{bol}$ and $R_{\rm BB}$, both peaking at $\sim-27$~d. A dense disk-like CSM funnels, and is partially engulfed by, \textit{Ejecta 1}. At $\sim-21$~d, \textit{Ejecta 2} is ejected with a velocity of $\sim5000$\kms (this could be the SN explosion) and almost immediately collides with \textit{Ejecta 1} wit some fast moving material escapes along less dense polar regions. $L_{bol}$ and $R_{\rm BB}$ follow the expansion of an opaque \textit{Ejecta 2} following the HV material seen in H$\alpha$. \textit{Ejecta 2} becomes optically thin and the photopshere begins to move inwards in velocity space. There is a linear decay in $R_{\rm BB}$ until $\sim+22$~d or the beginning of the plateau stage.$R_{\rm BB}$ plateaus at $\sim+25$~d due to effective internal heating from the site of interaction. Photons originating from the interaction site between \textit{Ejecta 1}, \textit{Ejecta 2} and the CSM begin to diffuse outwards, as the material becomes partially transparent. This coincides with the metamorphosis of the blue HV absorption to an emission profile.  At $\sim+45$~d, the \textit{Knee} stage drops in $L_{bol}$ and $R_{\rm BB}$ with $R_{\rm BB}$ at a slightly higher value when compared to the beginning of \eventB. Both red and blue emission lines narrow at this stage which may signify any intervening material is now completely optically thin and any escaping photons undergo minimal scattering. The dominant source of energy is now shock interaction due to ejecta-csm interaction. In the bottom left panel we also include a line of sight dependency. We expect the transient to show double-peaked emission lines when observed near the equator (e.g. \jbu, \bh, \al) and more singularly peaked when observed towards the polar regions (e.g. \ip). We note the similarities between this toy model and those proposed for \etacar \citep[e.g.][]{Smith2018}.}
\label{fig:geometry}
\end{figure*}

As discussed in Sect. \ref{sec:kinematices}, we proposed a double eruption model, where an first ejecta interacts with pre-existing CSM, followed by a second eruption some months later. The collision of these two ejecta produce the spectral and lightcurve evolution we present in \paperI and can be extrapolated to fit the observables of several \ip-like transients. We provide an illustration in Fig. \ref{fig:geometry} with a detailed outline of events in the provided caption.  

\subsection{Modelling the lightcurve using {\sc SNEC}}\label{sec:discuss_snec}

\begin{figure}
\includegraphics[width=\columnwidth]{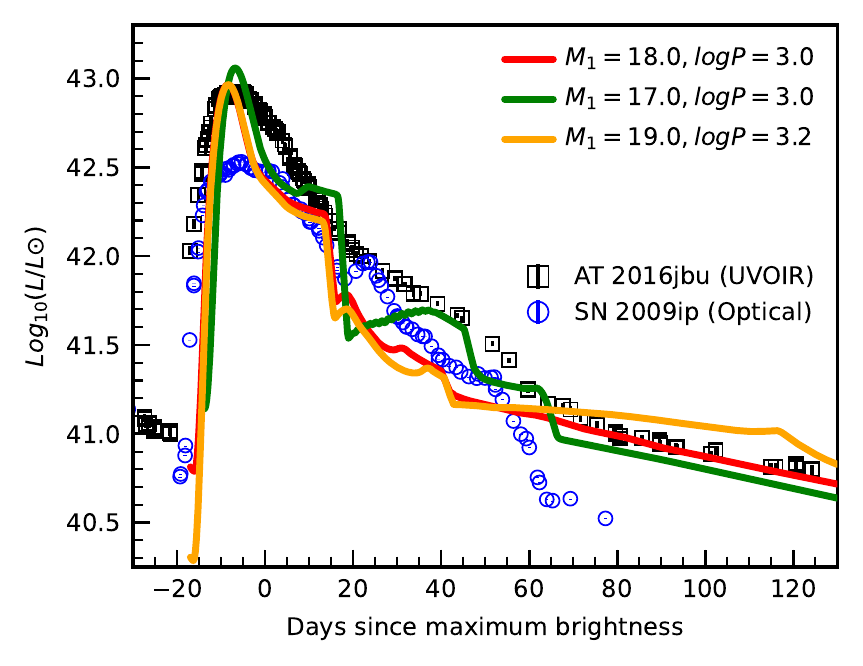}
\caption{Diagram showing the observed lightcurve and lightcurves simulated by {\sc SNEC} from progenitors which match the pre-explosion constraints. All include the circumstellar medium as described earlier. We include our pseudo-bolometric for \jbu in black and the optical pseudo-bolometric lightcurve for \ip  in blue\citep{Fraser_2013,Pastorello2013}.}
\label{fig:bpass+snec}
\end{figure}

To further explore the plausibility of the progenitor matching {\sc BPASS} models from Sect.~\ref{sec:discuss_progenitor}, we exploded a small subset of these with the SuperNova Explosion Code \citep[{\sc SNEC}, ][]{Morozova2015}. The full details of how {\sc BPASS} models are exported and exploded within {\sc SNEC} can be found in \citet{Eldridge2019}. The key addition to using the progenitor model structure is to add on a CSM component around the star. Here we use the values derived in Sect.~\ref{sec:kinematices} of a terminal wind velocity of 250~\kms and a mass-loss rate of 0.05~\msunperyr. For each of the input stellar models we use an explosion energy of $5.6\times10^{49}$~ergs, 0.016~\msun\ of $^{56}$Ni, and an inner mass cut at 5~\msun\ with nickel mixing out to 0.6~\msun. The resultant simulated bolometric lightcurves are shown in Fig.~\ref{fig:bpass+snec} and model parameters are given in Table \ref{tab:snec}.

\begin{table}
\centering
\caption{The parameters of the {\sc BPASS} models exploded with {\sc SNEC}.}
%Values of $M_2$ with an ``*'' indicate black hole companions. }
\label{tab:snec}
\begin{tabular}{ccccc}
\hline
$M_1$& $M_2$& & $M_{\rm final}$ & $M_{\rm CO }$\\
$/M_{\odot}$& $/M_{\odot}$& $\log (P_{\rm i}/{\rm days})$ & $/M_{\odot}$ & $/M_{\odot}$\\
\hline
17 & 11.9 &  3  & 5.9 &4.0 \\
% 17 & 13.6 &  3  &6.1 & 3.2 \\
18 & 16.2 &  3  & 6.5&  4.1 \\
% 19 &  1.9  & 0.6 & 7.3&  5.2 \\
19 & 13.3  & 3.2 &7.2&  5.4 \\
\hline
\end{tabular}
\end{table}

Our models are able to reproduce the magnitude of the peak luminosity, although exact matching of the lightcurve post-peak is difficult. Phases of the swept up wind becoming transparent, followed by the ejecta can be seen as the sudden drop-offs in Fig.~\ref{fig:bpass+snec}. We find that the width of the \eventB peak is dependent to some degree to how the density of the wind varies with distance from the star. The figure shows the resultant models where we assume $\rho_{\rm wind}(r) \propto r^{-1.6}$. We found that the shallower the density gradient the wider the peak and a best match is found with an exponent $n=-1.6$. 
In general, the models that best match the supernova lightcurve have low ejecta masses on the order of 1 to 2~\msun. Some models which have experienced a merger during their binary evolution have a higher ejecta mass do not match the lightcurve, being less luminous or evolving more slowly. Achieving an exact match between the models and observed lightcurve would require significant fine-tuning of the details of the CSM around the star, in terms of density profile, wind velocity, and the details of the wind acceleration. An exact match may also be impossible given the spherically symmetric assumptions of {\sc SNEC}. However, we take the reasonably close match between the model and observed lightcurves to indicate that a subset of the {\sc BPASS} models can explain \jbu. 

Intriguingly, the low $CO$ core mass of several of the progenitor models suggest an explosion close to the electron-capture regime where lower nickel masses and explosion energies would be expected.

%%%%%%%%%%%%%%%%%%%%%%%%%
%%%%%%%%% CCSNe %%%%%%%%%
%%%%%%%%%%%%%%%%%%%%%%%%%

\subsection{Was \jbu a Core-Collapse Supernova?}\label{sec:discuss_CCSN}

The main point of controversy is whether \jbu and \ip-like transients are indeed CCSNe; meaning the progenitor has been destroyed and the transient will eventually decay following a radioactive decay tail. This begs the question; If these are indeed CCSNe, \textit{when did core-collapse occur?}

\ip-like transients display two broad, luminous events, rather than the singularly peaked lightcurve, typically associated with SNe. \citet{Mauerhan_2013} suggest that \eventA is a CCSN and \eventB is a result of ejecta interacting with a dense CSM. In this scenario, with respect to \jbu, the duration of \eventA ($\sim$~60~days) is the time needed for this ejecta to reach the inner edge of the CSM. This scenario would be consistent with the early evolution of $R_{\rm BB}$ expanding at $\sim$700~\kms; however this velocity is implausibly slow for SN ejecta. More problematic still, at $-21$~d we see an increase in velocity where $R_{\rm BB}$ expands at $\sim$4500~\kms. In the case of core-collapse, we hence regard it as more plausible that \eventB is the terminal explosion of the progenitor, where the ejecta interacts with a non-terminal outburst that was ejected at $\sim700$~\kms\ around the start of \eventA. This scenario is also reinforced by the rise time ($\sim17$~days) and peak magnitude ($M_V\sim-18.5$~mag) of \eventB \citep{Nyholm2020}. 

We find a low value of $^{56}$Ni of $\lesssim$~0.016~\msun\ for \jbu, consistent with other \ip-like transients. Such a low $^{56}$Ni mass would be unusual for a normal CCSN, although an exception would be a faint electron capture SN (ECSN) or a sub-luminous Fe CCSN from a star with a ZAMS mass of around 8 -- 10~\msun. However, we find the mass of the \jbu progenitor to be significantly larger than that expected for an ECSN progenitor \citep{Doherty_2017}. Additionally the inferred explosion energy of $5.6\times10^{50}$~ergs (which may be a lower limit, as spherical symmetry is assumed) is too high for a typical ECSN \cite{Wanajo2009}. A final possibility that can explain such a low Ni mass (if this is a CCSN) is significant fallback onto a compact remnant \citep{Zampieri_1998,Benetti_2016}. 

Some challenges remain for the fallback scenario. A low metallicity environment is required, so that the progenitor star has retained much of its ZAMS mass \citep[e.g.][]{Heger2003}. This is hence an appealing scenario for \ip, due to its remote location \citep[$\sim$5~kpc from its host][]{Smith2016} and naturally low-metallicity environment. Conversely, this contradicts what we see for the environment around \jbu in Sect.~\ref{sec:muse}, where we find an approximate solar metallicity of 8.66~dex. It is hence expected that a $\sim$20~\msun\ progenitor will loose a significant fraction of its mass before exploding.

We see from Fig.~\ref{fig:muse_enviroment} that \jbu is located near a moderately star-forming region that is likely to host CCSNe, as seen from SN~1999ga. On the contrary, \ip is located on the outskirts of its host spiral galaxy, NGC~7259, at a galactocentric radius of $\sim$~5~kpc. \citet{Smith2016} finds no strong indication of massive star formation anywhere in the vicinity around \ip, unlike what is seen for \jbu. If the progenitor of \ip and \jbu are similar, as is suggested by their photometric and spectral evolution, then this begs the question why \ip is on its own. 

One of the biggest difficulties with \jbu as a CCSN is that it is in stark contrast to the predictions of single star stellar evolutionary models. A 20~\msun\ star is expected to end its life as a RSG which undergoes Fe core-collapse \citep{Heger2003}. From our {\sc dusty} modelling in Sect.~\ref{sec:progenitor}, we find that the progenitor of \jbu is not situated at the end of any single star evolutionary track. This suggests that the progenitor is not sufficiently evolved to undergo core-collapse. Our conclusion in Sect.~\ref{sec:progenitor} also suggests that the progenitor of \jbu is not a RSG but rather a YHG. We also note that if \jbu is indeed a CCSN, it is more appropriate to compare to the luminosity of the progenitor to the {\it terminal luminosity} of the models (typically corresponding to the end of core He-burning), in which case we find that it must have been a 12--16~\msun\ star. One must caution however that if the progenitor of \jbu was in a binary, then the expectations from single star evolution can be drastically altered. However, even if \jbu does arise from a binary progenitor system, models do not necessarily predict outbursts or eruptions immediately prior to explosion as seen in this case (discussed further in Sect. \ref{sec:discuss_binary} below). Clearly, further detailed stellar evolutionary modeling is required to fully explain the progenitor (or progenitor system) of \jbu.

A tantalising hint of a surviving progenitor is \jbu returning to its pre-explosion magnitude in 2019, as shown in Fig.~\ref{fig:hst_lc}. However, this detection may be serendipitous and further late time monitoring will be needed to confirm any surviving progenitor.

%%%%%%%%%%%%%%%%%%%%%%%%%%%%%%%%%
%%%%%%% Binary Modelling %%%%%%%%
%%%%%%%%%%%%%%%%%%%%%%%%%%%%%%%%%

\subsection{Binary Interaction}\label{sec:discuss_binary}

Several authors have suggested that \ip-like transients are a result of binary interaction \citep{smith13,Kashi2013,Soker_2013,Smith2018} as well as some other SN Impostors e.g. SN~2000ch \citep{Pastorello_2010,Smith_2011,Clark_2013}. Mass transfer within a binary system could naturally explain an asymmetric CSM environment, which we interpret as a circumstellar/circumbinary disk for \jbu.

\citet{Smith_2014} suggest that the isolated location of \ip may be explained as they are \textit{Kicked Mass Gainers} in a binary star system. For the progenitor to travel $\sim$~5~kpc within the required lifespan of a 50 -- 80~\msun\ star, a binary companion may be required to provide an additional source of fuel after the stars have been ejected \citep{Smith2016}. 

Binary merger events have recently been associated with Red Novae (RNe) and the more extreme, Luminous Red Novae (LRNe) \citep[see review by][and references therein]{Pastorello_2019}. These transients typically fall into the class of Gap Transients \citep{Kasliwal2012,PastorelloFraser_2019} and are amongst the most powerful stellar cataclysms. LRNe span a wide range of absolute magnitudes, from $-4$--$-15$~mag \citep{PastorelloFraser_2019}, and show a wide range of lightcurve shape and duration. 

The physical interpretation of LRNe is debated. The progenitors of LRNe are likely massive contact binaries, and the doubled peaked lightcurve is a consequence of a stellar merger plus a Common Envelope Ejection (CEE) \citep{Smith2016b,Metzger_2017,Pastorello_2019}. \citet{Pastorello_2019} suggest that there may be a continuum spanning between RNe to LRNe, with the possibility that this range can reach to brighter magnitudes (most likely caused by higher mass systems). \ip-like events may be some combination of a binary merger where the system consists of a relatively massive primary where the stars undergo a Common Envelope (CE) phase followed by a massive eruption. 

\jbu and \ip-like transients show a common peak magnitude and shape (i.e. \eventB appear to be similar among \ip-like events). We do not have adequate colour information for the peak of \eventA for \jbu however, \eventB has a colour of B-V$\sim$~0, which is comparable to that seen in LRNe in their first peak. \jbu never gets redder than $B-V$ of $\lesssim$~0.8 and after $\sim$~1.2 years, when the transient returns to a $B-V$ value of $\sim$~0.2. If \jbu is indeed related to LRNe, continued interaction in \jbu may be responsible for the relatively blue colour at late times.

\citet{Soker_2013} proposed that \ip is the result of the merger of a massive LBV with a binary companion in their ``mergerburst'' model. This model agrees with observations of \ip quite well, such as the moderate ejecta mass (a few \msun), most of which is moving at less than 5000~\kms. They further predict the remnant of their mergerburst will be a hot red giant star that will become apparent years after the transient fades, as is commonly associated with RNe and LRNe \citep[e.g][]{Pastorello_2019}. \citet{Kashi2013} discuss a similar explosion mechanism to the scenario we discuss in Sect.~\ref{sec:kinematices} and conclude the double-peaked lightcurve of \ip may be explained by to two successive outburst, separated by $\sim20$~days caused by periastron passages of the binary system.

It is appealing to conclude that \jbu is the result of a coalescing binary. This can naturally explain the historic variability, double-peaked lightcurve, and (inferred) asymmetric CS environment i.e. disk or bipolar outflow. \citet{Metzger_2017} proposed that LRNe can be well modeled by a single symmetric eruption in an asymmetric CSM environment. This asymmetric CSM is fueled by mass transfer within the binary over many orbits preceding the double-peaked event. The first peak of LRNe can be comfortably powered via cooling envelope emission from fast moving ejecta. Radiative shocks from the collision of this ejecta with material in the equatorial plane then power the second peak. This would be inconsistent with our proposed ``\textit{catch-up}" scenario for \jbu, although it cannot be ruled out conclusively.

We can speculate that the events prior to \eventB in \jbu and \ip-like events are similar to LRNe, including as mass transfer / Roche Lobe Overflow (RLOF) seen in the decade leading up to \eventA, and a merger/CEE powering \eventA itself. To explain the homogeneity of \eventB, the merging of the binary system must cause a violent (and possibly terminal) eruption.

Each \ip-like transient remains relatively blue for a long period of time, unlike what we see in LRNe, which is likely a sign of continued interaction. If we assume that \ip-like transients are indeed an upscaled version of LRNe, then this continued interaction at late times may reflect a more massive progenitor than is commonly associated with LRNe. In this scenario would expect a surviving star to become visible after this interaction has abated.

\section{Conclusion}\label{sec:conclusions}

In this paper, we have investigated the progenitor and environment of \jbu as well as modelling the transient itself. If \jbu is a single star, we find that the progenitor is consistent with a $\sim$~22~\msun\ progenitor \citep[e.g. Fig. 4 in][]{Smartt09}, with a color consistent with a YHG, roughly consistent with \citetalias{Kilpatrick2018}. Modelling of circumstellar dust using {\sc dusty} gives a luminosity and temperature of the progenitor similar to known YHGs. We show that the local environment around the progenitor of \jbu is consistent with a CCSN from a progenitor with ZAMS mass $\sim$~20~\msun; as the stellar population has an age of 15--200~Myr. We confidently rule out the possibility that the progenitor of \jbu is an LBV of 50--80~\msun, as has been proposed for \ip \citep{smith10,Foley2011}.

We find that the \textit{Event A/B} light curve can be modelled by two shells of material, with the later \eventB being powered by a ``catch-up’' scenario, involving two eruptive mass loss events and pre-existing CSM. Spectroscopic and photometric evolution is consistent with a spherically symmetric ejecta colliding with, and temporarily engulfing, previously ejected, asymmetric material. This interaction is the dominant energy source after $\sim$~2~months. After $\sim$ 200~days, \jbu shows increased interaction, likely reflecting a clumpy CSM.

\jbu shows tentative evidence for core-collapse. We find a upper limit of $^{56}$Ni of $\lesssim 0.016$~\msun\, but with strong on-going CSM interaction at this time, the real value of $^{56}$Ni is probably much lower (if any at all). Almost 1.5 years after maximum brightness, \jbu lacks signs of explosively nucleosynthesised material or emission from the burning products of late time stellar evolution.

We explore the possibility that \jbu is the result of a binary system. We compare our progenitor models with an extensive group of {\sc BPASS} models, exploring both CCSN and non-terminal events. We find that matching models have M$_{ZAMS}$ $\lesssim~26$~\msun. Steady state mass loss due to the progenitor wind is unable to produce the CSM density necessary to power the lightcurve and episodic mass loss may be required. Using {\sc SNEC} we find that a relatively low explosion energy (\SI{5.5e49}{ergs}) with a small ejecta mass ($\sim$~1-2~\msun) can comfortably power \jbu (assuming spherical symmetry). If we account for a high degree of asymmetry, we may have an explosion energy on par with a typical CCSN.

It appears that there is not a simple explanation for these transients. Following \textit{Hickam's dictum}, a low energy SN within a binary system with a disk-like CSM can account for the rise and peak of \eventB, low $^{56}$Ni, continued CSM interaction, and unique spectral features of \jbu. Additional binary interaction might explain \eventA e.g. due to a merger or CEE. Detailed modelling of this proposed scenario is beyond the scope of this paper and future work will involve exploring these scenarios in a non-symmetric setting.

The true nature of \jbu (and other \ip-like transients) remains elusive. Perhaps the ultimate answer will come if or when very late time observations reveal a surviving progenitor. To date, no conclusive evidence exists as to whether these transients destroy their progenitor. However, one must account for the possibility that if the progenitor survived, it may be obscured by a significant amount of dust. Deep images covering the full SED will hence be required to confidently rule out surviving, but dust-enshrouded, star. To this end, future observations with the upcoming {\it James Webb Space Telescope} will be essential. Alongside this, deep optical imaging from the Vera C. Rubin Observatory may capture similar pre-explosion variability in the years/decades prior to future \ip-like events, perhaps even allowing for a countdown timer before these events.

%%%%%%%%%%%%%%%%%%%%%%%%%%%%%%%%%%%%%%%%%%%%%%%%%%

\section*{Acknowledgements} 

S. J. Brennan acknowledges support from Science Foundation Ireland and the Royal Society (RS-EA/3471). M.F is supported by a Royal Society - Science Foundation Ireland University Research Fellowship. T.M.B was funded by the CONICYT PFCHA / DOCTORADOBECAS CHILE/2017-72180113 and acknowledges their financial support from the Spanish Ministerio de Ciencia e Innovaci\'on (MCIN), the Agencia Estatal de Investigaci\'on (AEI) 10.13039/501100011033 under the PID2020-115253GA-I00 HOSTFLOWS project, and from Centro Superior de Investigaciones Cient\'ificas (CSIC) under the PIE project 20215AT016. K.M. is funded by the EU H2020 ERC grant no. 758638. T.W.C acknowledges the EU Funding under Marie Sk\l{}odowska-Curie grant H2020-MSCA-IF-2018-842471, and thanks to Thomas Kr{\"u}hler for GROND data reduction. M.N is supported by a Royal Astronomical Society Research Fellowship. B.J.S is supported by NSF grants AST-1908952, AST-1920392, AST-1911074, and NASA award 80NSSC19K1717. M.S is supported by generous grants from Villum FONDEN (13261,28021) and by a project grant (8021-00170B) from the Independent Research Fund Denmark. L.H acknowledges support for Watcher from Science Foundation Ireland grant 07/RFP/PHYF295. Time domain research by D.J.S. is supported by NSF grants AST-1821987, 1813466, \& 1908972, and by the Heising-Simons Foundation under grant \#2020-1864. N.E.R. acknowledges support from MIUR, PRIN 2017 (grant 20179ZF5KS). Support for JLP is provided in part by ANID through the Fondecyt regular grant 1191038 and through the Millennium Science Initiative grant ICN12$\_$009, awarded to The Millennium Institute of Astrophysics, MAS. L.G. acknowledges financial support from the Spanish Ministry of Science, Innovation and Universities (MICIU) under the 2019 Ram\'on y Cajal program RYC2019-027683 and from the Spanish MICIU project PID2020-115253GA-I00. D.A.H and D.H are supported by AST-1911151, AST19-11225, and NASA Swift grant 80NSSC19K1639. G.P acknowledge support by the Ministry of Economy, Development, and Tourism’s Millennium Science Initiative through grant IC120009, awarded to The Millennium Institute of Astrophysics, MAS. L.T. acknowledges support from MIUR (PRIN 2017 grant 20179ZF5KS). Support for TW-SH was provided by NASA through the NASA Hubble Fellowship grant HST-HF2-51458.001-A awarded by the Space Telescope Science Institute, which is operated by the Association of Universities for Research in Astronomy, Inc., for NASA, under contract NAS5-26555. H.K. was funded by the Academy of Finland projects 324504 and 328898. 

This research made use of Astropy\footnote{\url{http://www.astropy.org}}, a community-developed core Python package for Astronomy \citep{astropy:2013, astropy:2018}. This research made use of data provided by Astrometry.net\footnote{\url{https://astrometry.net/use.html}}. Parts of this research were supported by the Australian Research Council Centre of Excellence for All Sky Astrophysics in 3 Dimensions (ASTRO 3D), through project number CE170100013. This research has made use of the NASA/IPAC Extragalactic Database (NED), which is operated by the Jet Propulsion Laboratory, California Institute of Technology, under contract with the National Aeronautics and Space Administration. We acknowledge Telescope Access Program (TAP) funded by the NAOC, CAS, and the Special Fund for Astronomy from the Ministry of Finance. This work was partially supported from Polish NCN grants: Harmonia No. 2018/30/M/ST9/00311 and Daina No. 2017/27/L/ST9/03221. This work made use of v2.2.1 of the Binary Population and Spectral Synthesis (BPASS) models as described in \citet{Eldridge2017} and \citet{Stanway2018}. This research is based on observations made with the NASA/ESA Hubble Space Telescope obtained from the Space Telescope Science Institute, which is operated by the Association of Universities for Research in Astronomy, Inc., under NASA contract NAS 5-26555. These observations are associated with program 15645. Observations were also obtained from the Hubble Legacy Archive, which is a collaboration between the Space Telescope Science Institute (STScI/NASA), the Space Telescope European Coordinating Facility (ST-ECF/ESAC/ESA) and the Canadian Astronomy Data Centre (CADC/NRC/CSA). This research has made use of the SVO Filter Profile Service\footnote{\url{http://svo2.cab.inta-csic.es/theory/fps/}} supported from the Spanish MINECO through grant AYA2017-84089.

%%%%%% APPENDIX %%%%%%%
\appendix

\section{Author Affiliations}
\label{app:affiliations}

$^{1}$  School of Physics, O’Brien Centre for Science North, University College Dublin, Belfield, Dublin 4, Ireland\\ 
$^{2}$  The Oskar Klein Centre, Department of Physics, AlbaNova, Stockholm University, SE-106 91 Stockholm, Sweden\\ 
$^{3}$  INAF-Osservatorio Astronomico di Padova, Vicolo dell’Osservatorio 5, I-35122 Padova, Italy\\ 
$^{4}$  Department of Physics and Astronomy, University of Turku, FI-20014, Turku, Finland\\ 
$^{5}$  The Department of Physics, The University of Auckland, Private Bag 92019, Auckland, New Zealand\\ 
$^{6}$  The Oskar Klein Centre, Department of Astronomy, AlbaNova, Stockholm University, SE-106 91 Stockholm, Sweden\\ 
$^{7}$  Max-Planck-Institut f\"{u}r Extraterrestrische Physik, Giessenbachstra\ss e 1, 85748 Garching, Germany\\ 
$^{8}$  Department of Astronomy, The Ohio State University, 140 W. 18th Avenue, Columbus, OH 43210, USA\\ 
$^{9}$  Center for Cosmology and AstroParticle Physics (CCAPP), The Ohio State University, 191 W. Woodruff Avenue, Columbus, OH 43210, USA\\ 
$^{10}$ Department of Physics and Astronomy, Texas A$\&$M University, 4242 TAMU, College Station, TX 77843, USA\\ 
$^{11}$ Cerro Tololo Inter-American Observatory, NSF’s National Optical-Infrared Astronomy Research Laboratory, Casilla 603, La Serena, Chile\\ 
$^{12}$ Institut d’Astrophysique de Paris (IAP), CNRS $\&$ Sorbonne Universite, France\\ 
$^{13}$ Kavli Institute for Astronomy and Astrophysics, Peking University, Yi He Yuan Road 5, Hai Dian District, Beijing 100871, China\\ 
$^{14}$ NINAF-Osservatorio Astronomico di Padova, Vicolo dell’Osservatorio 5, I-35122 Padova, Italy\\ 
$^{15}$ Institute of Space Sciences (ICE, CSIC), Campus UAB, Carrer de Can Magrans s/n, 08193 Barcelona, Spain\\ 
$^{16}$ Steward Observatory, University of Arizona, 933 North Cherry Avenue, Tucson, AZ 85721-0065, USA\\ 
$^{17}$ Department of Physics, Florida State University, 77 Chieftan Way, Tallahassee, FL 32306, USA\\ 
$^{18}$ Tuorla Observatory, Department of Physics and Astronomy, FI-20014 University of Turku, Finland.\\ 
$^{19}$ Finnish Centre for Astronomy with ESO (FINCA), FI-20014 University of Turku, Finland\\ 
$^{20}$ CBA Kleinkaroo, Calitzdorp, South Africa\\ 
$^{21}$ Departamento de Ciencias F\'isicas, Universidad Andres Bello, Avda. Republica 252, Santiago, 8320000, Chile\\ 
$^{22}$ Millennium Institute of Astrophysics, Santiago, Chile\\ 
$^{23}$ Department of Astronomy/Steward Observatory, 933 North Cherry Avenue, Rm. N204, Tucson, AZ 85721-0065, USA\\ 
$^{24}$ Institute for Astronomy, University of Hawai’i, 2680 Woodlawn Drive, Honolulu, HI 96822, USA\\ 
$^{25}$ Astrophysics Research Centre, School of Maths and Physics, Queen’s University Belfast, Belfast BT7 1NN, UK\\ 
$^{26}$ Mt Stromlo Observatory, The Research School of Astronomy and Astrophysics, Australian National University, ACT 2601, Australia\\ 
$^{27}$ National Centre for the Public Awareness of Science, Australian National University, Canberra, ACT 2611, Australia\\ 
$^{28}$ The ARC Centre of Excellence for All-Sky Astrophysics in 3 Dimension (ASTRO 3D), Australia\\ 
$^{29}$ Astronomical Observatory, University of Warsaw, Al. Ujazdowskie 4, 00-478 Warszawa, Poland\\ 
$^{30}$ Institute of Astronomy, Madingley Road, Cambridge,CB3 0HA, UK\\ 
$^{31}$ Unidad Mixta Internacional Franco-Chilena de Astronom\'ia, CNRS/INSU UMI 3386 and Instituto de Astrof\'isica, Pontificia Universidad Cat\'olica de Chile, Santiago, Chile\\ 
$^{32}$ Aix Marseille Univ, CNRS, CNES, LAM, Marseille, France\\ 
$^{33}$ RHEA Group for ESA, European Space Astronomy Centre (ESAC-ESA), Madrid, Spain\\ 
$^{34}$ nstitute of Space Sciences (ICE, CSIC), Campus UAB, Carrer de Can Magrans, s/n, E-08193 Barcelona, Spain.\\ 
$^{35}$ Tuorla Observatory, Department of Physics and Astronomy, FI-20014 University of Turku, Finland\\ 
$^{36}$ Kavli Institute for Cosmology, Institute of Astronomy, Madingley Road, Cambridge, CB3 0HA, UK\\ 
$^{37}$ Las Cumbres Observatory, 6740 Cortona Drive, Suite 102, Goleta, CA 93117-5575, USA\\ 
$^{38}$ Department of Physics, University of California, Santa Barbara, CA 93106-9530, USA\\ 
$^{39}$ The Observatories of the Carnegie Institution for Science, 813 Santa Barbara St., Pasadena, CA 91101, USA\\ 
$^{40}$ School of Physics $\&$ Astronomy, Cardiff University, Queens Buildings, The Parade, Cardiff, CF24 3AA, UK\\ 
$^{41}$ School of Physics and Astronomy, University of Southampton, Southampton, Hampshire, SO17 1BJ, UK\\ 
$^{42}$ School of Physics, Trinity College Dublin, The University of Dublin, Dublin 2, Ireland\\ 
$^{43}$ Department of Physics, University of the Free State, PO Box 339, Bloemfontein 9300, South Africa\\ 
$^{44}$ Carnegie Observatories, Las Campanas Observatory, Colina El Pino, Casilla 601, Chile\\ 
$^{45}$ Birmingham Institute for Gravitational Wave Astronomy and School of Physics and Astronomy, University of Birmingham, Birmingham B15 2TT, UK\\ 
$^{46}$ Institute for Astronomy, University of Edinburgh, Royal Observatory, Blackford Hill, EH9 3HJ, UK\\ 
$^{47}$ Nucleo de Astronomıa de la Facultad de Ingenierıa y Ciencias, Universidad Diego Portales, Av. Ej´ercito 441, Santiago, Chile\\ 
$^{48}$ Department of Physics and Astronomy University of North Carolina at Chapel Hill Chapel Hill, NC 27599, USA\\ 
$^{49}$ Department of Physics, Florida State University, 77 Chieftain Way, Tallahassee, FL 32306-4350, USA\\ 
$^{50}$ Department of Physics and Astronomy, Aarhus University, Ny Munkegade, DK-8000 Aarhus C, Denmark\\ 
$^{51}$ Department of Physics, University of Warwick, Coventry CV4 7 AL, UK\\ 
$^{52}$ Astrophysics Research Centre, School of Mathematics and Physics, Queen`s University Belfast, Belfast BT7 1NN, UK\\ 
$^{53}$ Institute of Space Sciences (ICE, CSIC), Campus UAB, Carrer de Can Magrans, s/n, E-08193 Barcelona, Spain.
$^{54}$ Center for Astrophysics \textbar{} Harvard \& Smithsonian, 60 Garden Street, Cambridge, MA 02138-1516, USA
$^{55}$ The NSF AI Institute for Artificial Intelligence and Fundamental Interactions

\section*{Data availability}

The photometric and spectroscopic data underlying this article are available as described in Paper I. The BPASS models are available at \url{https://bpass.auckland.ac.nz/}, while HST data are available at the Mikulski Archive for Space Telescopes \url{at https://archive.stsci.edu}.
%%%%%%%%%%%%%%%%%%%%%%%%%%%%%%%%%%%%%%%%%%%%%%%%%%


\bibliographystyle{mnras}
\bibliography{AT2016jbu_Paper_II_references} 

% Don't change these lines
\bsp	% Typesetting comment
\label{lastpage}
\end{document}